\documentclass{bcs_computer_updated/comjnl}

\usepackage{algorithm2e}
\usepackage[utf8]{inputenc}
\usepackage[english]{babel}
\usepackage[table]{xcolor}
\usepackage[T1]{fontenc}    
\usepackage{url}            
\usepackage{booktabs}
\usepackage{tabularx} 
\usepackage{amsfonts}
\usepackage{nicefrac}
\usepackage{microtype}
\usepackage{graphicx}
\usepackage[rightcaption]{sidecap}
\usepackage{amsmath} 
\usepackage{pgfplots}
\usepackage[super]{nth}
\usepackage{todonotes}
\usepackage{multirow}
\usepackage{pgf}
\usepackage{import}
\usepackage{amsmath}

\usepackage{enumitem}
\usepackage{wrapfig}
\usepackage{multirow}
\usepackage{siunitx}
\usepackage{etoolbox}
\usepackage[colorlinks=false,citecolor=blue]{hyperref}
\usepackage{caption}
\usepackage{subcaption}
\usepackage[title]{appendix}
\usepackage{float}
\usepackage{cleveref}
\usepackage{svg}

\pgfplotsset{compat=1.17}

\RestyleAlgo{ruled}
\SetKwInOut{Parameter}{Parameters}
\SetKwComment{Comment}{$\triangleright$\ }{}
\SetKwProg{Pn}{Function}{:}{\KwRet}

\newcommand{\removelatexerror}{\let\@latex@error\@gobble}

\newcommand{\orcid}[1]{\href{https://orcid.org/#1}{\includegraphics[width=10pt]{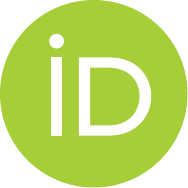}}}

\title[Near Real-Time Social Distance Estimation in London]{Near Real-Time Social Distance\\Estimation in London}

\author{James Walsh\textsuperscript{1}\orcid{0000-0002-2100-7696}}
\author{Oluwafunmilola Kesa\textsuperscript{2}\orcid{0000-0002-0609-9377}}
\author{Andrew Wang\textsuperscript{4}\orcid{0000-0003-0838-7986}}
\author{\\Mihai Ilas\textsuperscript{4}\orcid{0000-0003-4270-9846}}
\author{Patrick O'Hara\textsuperscript{2}\orcid{0000-0001-9600-7554}}
\author{Oscar Giles\textsuperscript{1}\orcid{0000-0002-4056-1916}}
\author{Neil Dhir\textsuperscript{1,2}\orcid{0000-0002-9193-8824}}
\author{Mark Girolami\textsuperscript{1,5}\orcid{0000-0003-3008-253X}}
\author{Theodoros Damoulas\textsuperscript{1,2,3}\orcid{0000-0002-7172-4829}}
\affiliation{\textsuperscript{1}The Alan Turing Institute; Departments of \textsuperscript{2}Computer Science and \textsuperscript{3}Statistics,\\University of Warwick; \textsuperscript{4}Christ's College, University of Cambridge;\\
\textsuperscript{5}Department of Engineering, University of Cambridge} \email{ \{jwalsh, ogiles, ndhir\}@turing.ac.uk; \{funmi.kesa, patrick.h.o-hara, t.damoulas\}@warwick.ac.uk; \{aslw3, mai32\}@cantab.ac.uk; mag92@cam.ac.uk } 

\shortauthors{J. Walsh, O. Kesa, A. Wang, M. Ilas, P. O'Hara, O. Giles, N. Dhir, M. Girolami, T. Damoulas}

\received{Day Month 2022}


\begin{document}

\begin{abstract}
To mitigate the current COVID-19 pandemic, policy-makers at the Greater London Authority, the regional governance body of London, UK, are reliant upon prompt, accurate and actionable estimations of lockdown and social distancing policy adherence. Transport for London, the local transportation department, reports they implemented over 700 interventions such as greater signage and expansion of pedestrian zoning at the height of the pandemic’s first wave with our platform providing key data for those decisions. Large well-defined heterogeneous compositions of pedestrian footfall and physical proximity are difficult to acquire, yet necessary to monitor city-wide activity (``busyness'') and consequently discern actionable policy decisions. To meet this challenge, we leverage our existing large-scale data processing urban air quality machine learning infrastructure to process over 900 camera feeds in near real-time to generate new estimates of social distancing adherence, group detection and camera stability. In this work we describe our development and deployment of a computer vision and machine learning pipeline. It provides near immediate sampling and contextualisation of activity and physical distancing on the streets of London via live traffic camera feeds. We introduce a platform for inspecting, calibrating and improving upon existing methods, describe the active deployment on real-time feeds and provide analysis over an 18 month period.
\end{abstract}

\maketitle

\section{Introduction}\label{section_intro}
Before 2020, the phrase ``social distancing'' had hardly any visibility to the public eye \cite{socdist_googletrends} as vernacular more frequently found in epidemiology textbooks and historical reports \cite{socdist_googlebooks}. However, during the COVID-19 pandemic, physical spacing between strangers became a means of trying to curb the spread of the virus.

As the global community is actively engaged in understanding more about the effects and transmission mechanisms of COVID-19, many governments have enacted temporary restrictions targeted at reducing the proximity of the public to one another. Measures such as limiting capacity within enclosed spaces, communicating new pedestrian traffic flow, and when necessary, enacting broader controls via ``lock-downs'' \cite{may_lockdown-type_2020}. The monitoring of public response to these measures have come out of necessity for policy makers to better understand their adoption, plan economic recovery and eventual suspension. When social restrictions were first implemented in the UK, there were limited measures of public activity in the context of likely vectors for viral transmission. A number of private companies trading in public movement data began providing aggregate information at the request of local government, from sources such as workplace reporting, wearable sports activity trackers and point of sale transactions \cite{gla_covid_mobility}. It became clear there was an immediate need for additional response metrics for pedestrian activity, unmet by the aforementioned sources.

\begin{figure*}[t]
\centering
    \begin{subfigure}[t]{0.98\columnwidth}
        \centering
        \includegraphics[width=\textwidth]{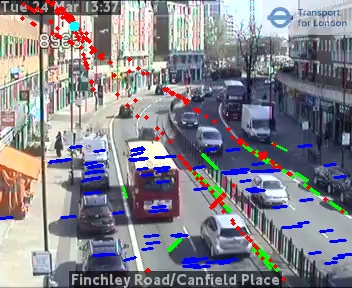}
        \caption{Road curvature challenging condition example, leading to erroneous vanishing point.}
    \end{subfigure}
    \hspace{2pc}
    \begin{subfigure}[t]{0.98\columnwidth}
        \centering
        \includegraphics[width=\textwidth]{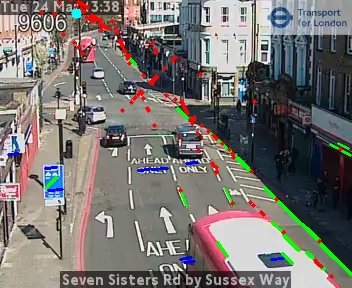}
        \caption{Harsh shadow parallel to vanishing lines, a beneficial scenario.}
    \end{subfigure}
    \caption{Lines detected by our feature extraction algorithm; two orthogonal sets of lines: those parallel to the foreground road (green, \emph{road edges}) and those perpendicular (blue, \emph{road perpendiculars}). The intersection of each set, the vanishing point (light blue) which lies on the horizon. Some challenging conditions are visible, including varying lighting and non-zero road curvature.\label{fig:edges}}
    \vspace{-15pt}
\end{figure*}

This work seeks to estimate social distance in areas of high footfall in London, UK. The goal is to gauge adherence at high spatial and temporal granularity, and most importantly, provide near real-time access to policy makers. We describe a \textit{social-distancing estimation system} using Open Government Licensed \cite{ogl} traffic cameras directed towards pedestrian crossings and pavements. We include the description of our pipeline, methodology, algorithms and new accuracy results as urban object detection benchmarks.

The basis for this approach was initially built for constructing greater predictive features to improve a live air quality model of London \cite{hamelijnck2019multiresolution}. It is known that pollutants are generated at different rates depending on driving activity \cite{BIGAZZI2012538}. Traffic camera footage is a suitable candidate for proving features on typical vehicle movement, capable of assisting the modelling of fine airborne particulates contributing to air pollution. The cloud infrastructure developed for the air quality model serves as the foundation for our social distancing estimation system.

Due to the nature of large-scale CCTV capture, there were initial substantial privacy concerns. All footage employed throughout the process is anonymised via deliberate restrictive sampling and systematically undergoes continuous review by The Alan Turing Institute’s Ethical Advisory Group \cite{turing_eag}.

\begin{figure*}[t]
    \centering
    \begin{subfigure}[t]{0.98\columnwidth}
        \centering
        \includegraphics[width=\textwidth]{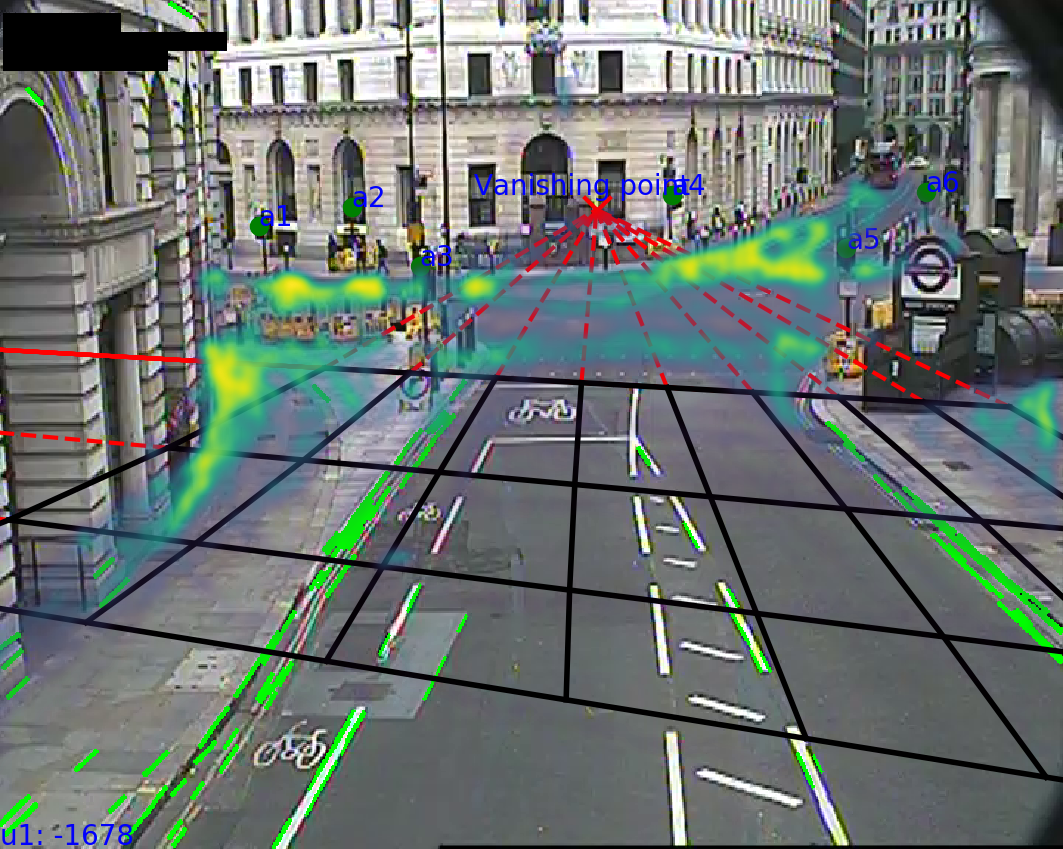}
        \caption{Detected edges, generated ground plane, and overlaid pedestrian detection density in bright green, black, and viridis heat map respectively.\label{fig:lines.a}}
    \end{subfigure}
    \hspace{2pc}
    \begin{subfigure}[t]{0.98\columnwidth}
        \centering
        \includegraphics[width=\textwidth]{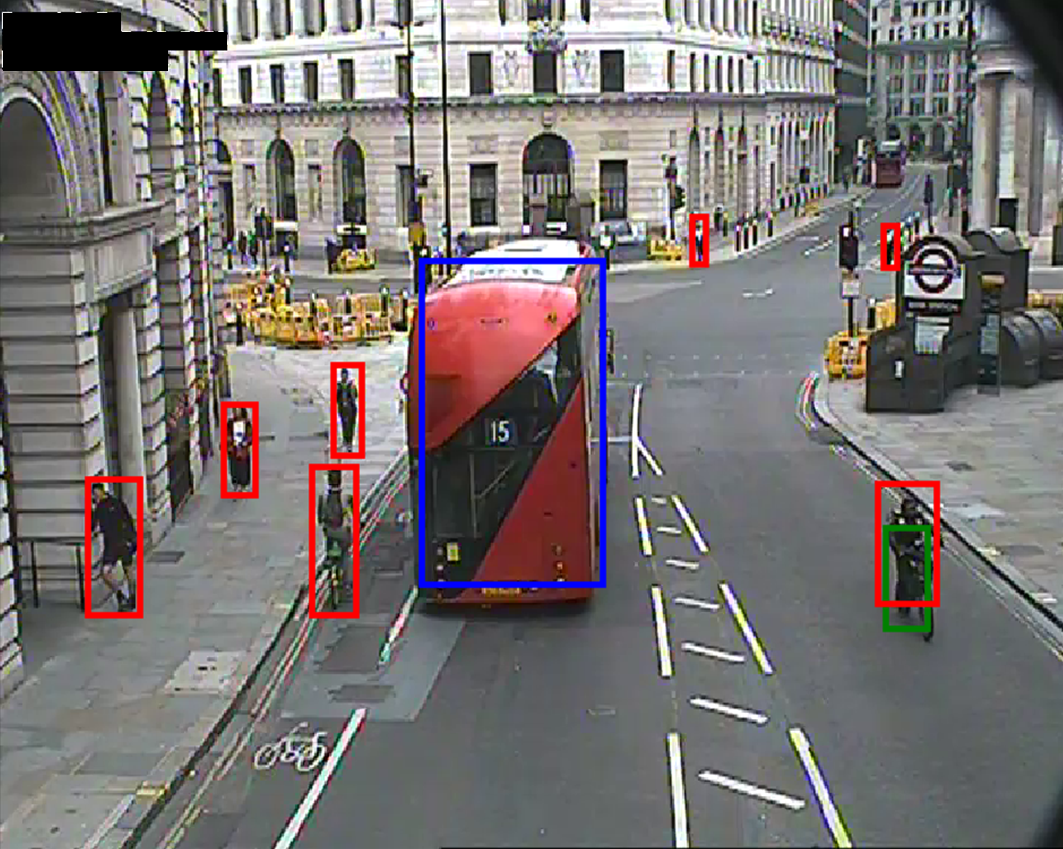}
        \caption{High accuracy detections, pedestrians, buses and bicycles in red, blue, green respectively. \label{fig:lines.b}}
    \end{subfigure}
    \caption{Example of our method applied to a traffic camera at Bank, London\label{fig:lines}.}
    \vspace{-15pt}
\end{figure*}

\section{Method}\label{section_methods}
Cameras available to the public are heterogeneous in quality and fall victim to the sporadic physical nature of London's historical streets. This scenario presents numerous challenges from a geospatial statistical and technical perspective, see \cref{fig:lines} for an example of a post-processed still frame. Our platform predominately relies upon 912 independent traffic camera feeds, over 500 of which typically overlook an intersection or crossing with an expected pedestrian footfall. In order to mitigate potential deanonymisation, all input visual data is reduced in video resolution, significantly hampering facial identification. Additionally, to place our results in an appropriate broader context, our research goal is to measure variations in social distancing and feed quality over extended periods of time.

Each camera feed provides two data elements: a short video every few minutes and a restrictive set of static metadata regarding location and approximate cardinal direction. Hence, before attempting to estimate any pedestrian location, each camera requires an initial digital twin (DT) abstraction to define the \textit{world-plane} of the visible scene stage, usually synonymous with visible road structure. A final \textit{real-world} calibration is applied using human-labelled mappings from pixel locations within the image to physical coordinates of objects identified within the scene. These \textit{anchors} are considered as ``ground-truth,'' examples include road markings, telephone booths and traffic lights. The objects selected are collectively referred to as ``urban furniture,'' and are of most benefit if visible from aerial or satellite photography for later calibration. As a form of \textit{image registration}, this enables mapping from the two-dimensional video frame to an inferred unreferenced world-plane, finally to a real-world location. The process is difficult with highly variate CCTV scenes; our method learns one set of parameters for mapping a 2D scene to a 3D real-world coordinate projection and is described in \cref{sect:cam_cal}.

Once complete, active data collection continuously ingests ten-second camera clips from the public domain. Upon successful retrieval of each video sample, they are batched for object detection. Our image processing pipeline is composed of a cluster of dynamically-scaled compute resource via virtual machines operated by a container-orchestration system called Kubernetes \cite{k8s}. A batch of 500 clips are ingested at a time, we are careful to ensure our model and computational resources are sufficient to process each sample in less or equal time than they represent, i.e. 30 minutes of footage must complete within 30 minutes, or the system would perpetually slip behind real-time. We employ a tuned state-of-the-art object detection model called YOLOv4 \cite{bochkovskiy2020yolov4, yolov4github} for identification of pedestrians within video frames. The reasoning for this selection and the tuning process is included in \cref{sect:obj_dect}, and active deployment as described in \cref{sect:deploy}.

Results from the camera calibration and object detection stages are then stored within high-availability databases \cite{postgres} near our data storage and image processing cluster. These databases permit immediate availability to public policy makers, specifically the greater London authority (GLA) and transport for London (TFL) via a reliable representational state transfer application programming interface (REST API). Additionally, high availability increases capacity for complimentary research tasks, such as simultaneously watching for spikes or irregularities via expectation-based network scan statistics \cite{haycock2020expectationbased}.

A primary challenge borne out of long-term experimental processing is the unexpected consequences of relying upon cameras prone to real-world interference. Some examples identified during the developmental phase of this system are graffiti, wind progressively drifting the view direction, physical malfunction, and trees sprouting leaves restricting previously clear views. In response to these detriments, we designed a camera stability change point detection process for identifying and alerting when scene dissimilarity meets a predetermined threshold, as described in \cref{sect:sscd}.

Finally, purely recognition, localisation and relative distance are not enough for adequate social distancing metrics, as pedestrian activity typically includes grouping behaviour. As individuals seek to preserve physical distance with strangers whilst reducing the chance of disbanding their safe social group (or ``bubble'') \cite{bubbles}; hence, an inclusion of group agency is considered. An algorithm for group detection operating at frame and scene granularity is presented in  \cref{sect:group_detect} for discussion as part of the final results.

\vspace{-5pt}
\subsection{Camera Calibration\label{sect:cam_cal}}

Obtaining a \textit{world-plane mapping} of a camera scene is extensively described in the computer vision and photogrammetry literature \cite{Dewitt2000ElementsOP, Zitov2003ImageRM}. A large portion of literature requires manual calibration using known patterns to estimate a mapping from sensor data to a real world contextualisation \cite{caprile, faugeras, tsai}. We aim to learn the geometric relationship between camera view and physical scene, frequently described through similarity, affine or projective transformations. A \textit{vanishing point} of an image is the location of apparent three-dimensional convergence of parallel \textit{vanishing lines} from a two-dimensional perspective (\cref{fig:lines.a}). Estimation of these vanishing lines is a common technique to recover some of these transformations. Our input scenes have multiple limitations: roads are usually curved or contain junctions of varying width; irregular road markings vary in quality; low video resolution of the feed; lighting conditions change frequently and are individually very short in duration. 

Scene object context methods \cite{schoepflin,dubska} use the activity of multiple vehicles travelling parallel and regularly to estimate the vanishing point. Which is feasible solution for our problem if multiple samples were stitched together and vehicle movement manually corrected. Calibration methods \cite{lai,song,dong,fung} require clear, regular or known lines in the scene, which is not practical in the case of a large spread of physical geometry. A stratified transformation approach discussed in \cite{liebowitz} relies upon maximum likelihood estimation (MLE), a popular method for parameter estimation of an assumed probability distribution given some observations. This is applied over multiple extracted lines from high quality images to build a real-world model, an issue for our low resolution samples. Finally, \cite{schoepflin, song, dong, cathey} extract visible road features by using a derivative-based binarisation operator. This is principally suitable for cameras overlooking straight and visually similar lanes, which is turn is only suitable for a portion of our input domain. Overall, we sought a more easily generalisable method considerate of our cluttered urban traffic scenes at low resolution that leverages our high sample quantity.
\vspace{-5pt}
\subsubsection{Simplified Pinhole Camera Model}
A mapping, $(u,v,0) \mapsto (X,Y,Z)$, is sought from the image-plane to world geometry -- for example, the transformation from pixels representing the bikes in \cref{fig:lines} to a physical location. Without a priori truth of any parameters describing the camera properties, these properties should be estimated or assumed and categorised into two groups: \textit{intrinsics} and \textit{extrinsics}. Examples of intrinsics include focal length, principal point, skew and aspect ratio, whereas extrinsics include positioning and direction. After manual inspection of all cameras, we conclude the suitability of the \textit{Simplified Pinhole Camera Model}, as fewer than 0.5\% of cameras have ultra wide-angle (``fisheye'') lenses.

\begin{figure*}[t]
    \centering
    \begin{subfigure}[t]{0.98\columnwidth}
        \centering
        \subfloat{\resizebox{0.98\columnwidth}{!}{\input{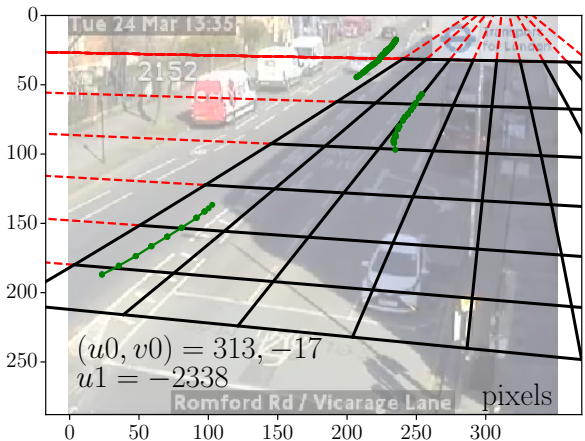}}}
        \caption{Intermediate perspective mapping from image plane. \label{fig:gridtracking1}}
    \end{subfigure}
    \hspace{2pc}
    \begin{subfigure}[t]{0.98\columnwidth}
        \centering
        \subfloat{\resizebox{0.98\columnwidth}{!}{\input{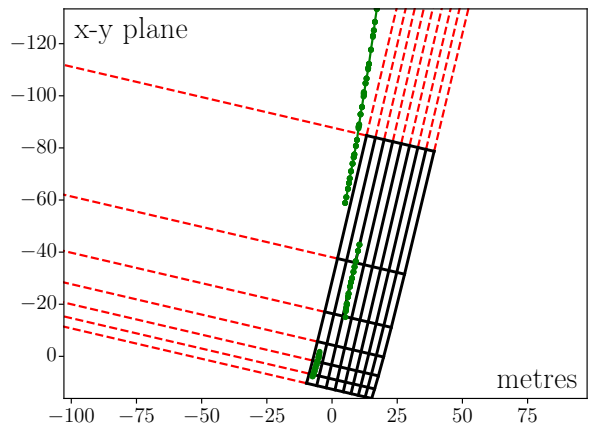}}}
        \caption{Intermediate world plane from perspective mapping. \label{fig:gridtracking2}}
    \end{subfigure}
    \caption{Demonstration of perspective mapping of camera calibration from image to world plane before estimated registration to British National Grid. Rays (black, solid) are drawn as grid lines and extended (red, dashed) to the estimated vanishing points $(u_0,v_0)$ and $(u_1,v_0)$. After mapping onto world coordinates, for example, vehicle trajectories (green, dotted) are also mapped by this transformation.\label{fig:gridtracking}}
    \vspace{-15pt}
\end{figure*}

Our simplified pinhole camera model allows the transformation to be described by four parameters $u_0,v_0,u_1,h$ where $(u_0,v_0),(u_1,v_0)$ are the vanishing points of two orthogonal planar directions subtending the horizon line, and $h$ is the height of the camera above ground (\cref{fig:gridtracking}). Parallel lines on the road and on cars, such as road edges, advanced stop lines and car and truck edges, are used to estimate this transformation (\cref{fig:lines.a}). This model makes the following assumptions:
\begin{enumerate}[label=(\alph*), noitemsep, nolistsep]
    \item Unit skew, i.e. regularly square pixel grid.
    \item Constant aspect ratio, i.e. no change in width to height ratio of pixels.
    \item Coincidence of principal point and image centre, i.e. no change in the center pixel from the center of the camera view.
\end{enumerate}
These are commonplace and rarely estimated when lacking more detailed visual information \cite{criminisi}, \cite{dubska}. External assumptions as follows:
\begin{enumerate}[label=(\alph*), noitemsep, nolistsep,start=4]
    \item Rectilinear lens, i.e. zero radial distortion; the image has already been pre-corrected such that perpendicular straight lines in reality are straight on the perpendicular pixel grid.
    \item Flat horizon $v_0=v_1$, i.e. camera has zero-roll.
    \item Zero-inclined roads $Z=0$, i.e. pedestrians do not move in a space large enough to calibrate deviation in elevation.
\end{enumerate}
Where cameras fail these external assumptions, a pre-processing stage included additional information to correct radial distortion \cite{dubska}, inclined horizon (setting $v_1 \neq v_0$), and non-zero inclination $Z$ \cite{schoepflin2}.

\vspace{-5pt}
\subsubsection{Edge Detection}
Our method for edge detection should be robust in noisy, low resolution scenes with varying light conditions. Whilst deep learning approaches for edge detection such as Visual Geometry Group ("VGG") models \cite{SimonyanZ14a} have seen significant advancements in the last decade \cite{7410521, 8100105}, they require hours of training on a large set of labelled edges. Note our dataset does not have labelled edges. To produce these data for this task outweighs the value of rapid perspective mapping on 912 scene samples.
This problem extends to considering direct vanishing point estimation. The aforementioned deep learning approaches would require labelled data on the order of magnitude of hundreds of samples \cite{8100105} for direct perspective estimation. We instead turn our attention to classical methods \cite{shapiro}.

Developed in 1986, the Canny Edge Detector has seen wide adoption for its adept ability to find edges under the edge detection goals of low error rates and minimised false edges in noisy-scenarios, suitable for our low resolution highly light-variate input scenes, see Figures \ref{fig:edges} and \ref{fig:lines.a}.

The method relies upon \textit{Gaussian filters} to first smooth potential noise and then applies four filters to find \textit{intensity gradients} with reference to gradient angle direction. Edge-thinning is subsequently applied via \textit{magnitude thresholding}, but this is not enough to remove spurious variations in colour and noise. A second \textit{double threshold} is applied using the surviving edge gradients, this time utilising high and low empirically determined from the whole edge set. Finally, some final weak edges remain. A process called \textit{hysteresis} is applied via blob analysis to determine survival based on proximity to neighbouring strong pixels. 

\vspace{-5pt}
\subsubsection{Parameter Estimation}
In order to learn our simplified  pinhole  camera  model detector parameters $(u_0, v_0, u_1, h)$, scenes with light vehicle traffic are selected and the edge detector applied per frame to find sets of \emph{road edges} and \emph{road perpendiculars} as shown in \cref{fig:lines}. In order to learn a set of vanishing lines from these edges, the Hough transform matches collinear edge segments into linked lines which are then filtered by gradient \cite{schoepflin} and dimensions. The vanishing point is then simply estimated as the highest frequency of the pairwise line intersections. This is chosen over more computationally expensive aforementioned MLE methods \cite{zhang} where the vanishing point error is optimised using least squares \cite{criminisi}, \cite{cathey} or Levenberg–Marquardt \cite{schoepflin}, \cite{liebowitz}. 
This procedure is repeated across different contrast factors to provide a robust line detector in challenging lighting conditions that such as British weather traditionally exhibits. These $u_0,u_1,v_0$ values are averaged over all frames to extract a final estimate.

Finally, the camera height $h$ must be manually estimated. One method is by transforming an object of known dimensions. For example, using frequently appearing London buses of fixed 4.95m height, the calculated height averages $h = 9.6m$ with 10\% average deviation across 7 randomly picked cameras. Other ways to obtain the scale $h$ include using car length averages \cite{schoepflin2} or known lane spacing \cite{fung,cathey}. Given few known consistent standardised urban furniture upright heights, the London Bus method is appropriate. With each parameter estimated is it possible to define a world-plane.\\
\newline

\subsubsection{Real-world Reference}
\begin{figure}[t]
\centering
\vspace{3pt}
\includegraphics[width=\columnwidth]{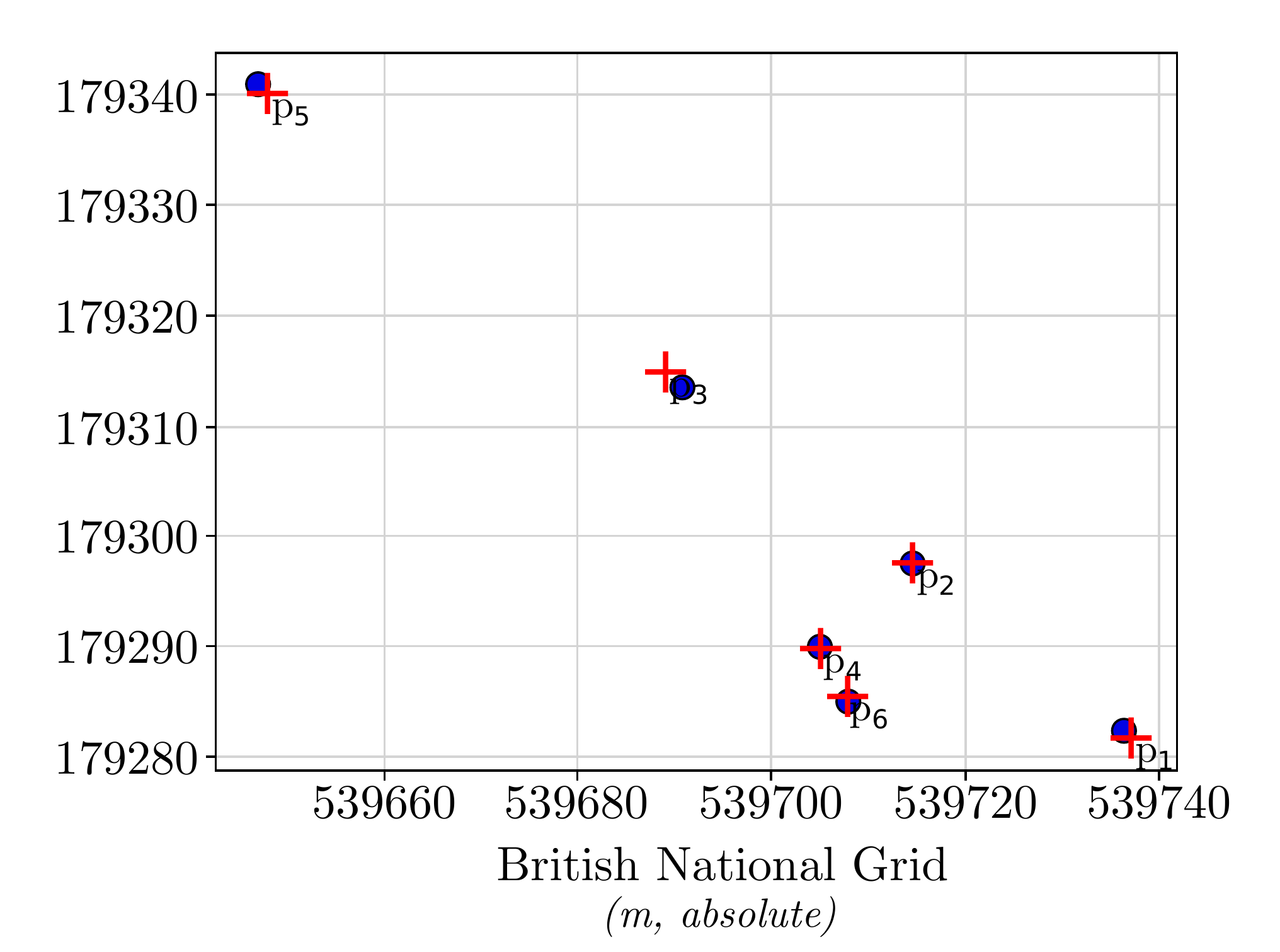}

\vspace{-10pt}
\caption{Estimated locations of urban furniture (green) and transformed ground-truth scene anchors (blue) on British National Grid.\label{fig:bns_cal}}
\vspace{-5pt}
\end{figure}

The world-plane projection (\cref{fig:gridtracking}) is as yet unreferenced to the real-world; a Euclidean Norm may be applied but not uniformly across all cameras. We employ points of reference via \textit{geotagged} static urban furniture,. such as traffic lights or road markings to map this intermediate world-plane to a real-world representation,

$$ \begin{bmatrix} x'+e_x \\ y'+e_y \\ 0 \end{bmatrix} = \begin{bmatrix} k_1 \cos(\theta) & k_3 \sin(\theta) & t_x \\ -k_2 \sin(\theta) & k_4 \cos(\theta) & t_y \\ 0 & 0 & 1 \end{bmatrix} \begin{bmatrix} x \\ y \\ 0 \end{bmatrix} $$

We select an appropriate flat 2D projected coordinate reference system, British National Grid (OS 27700). We then employ a second transformation between these two 2D Cartesian frames of reference, represented above with scale and shear factors, $k$, angle of rotation, $\theta$, translations $t$, and error terms, $e$. The estimated real-world representation is the result of optimisation of the sum of squares error between transformed image-coordinates of the urban furniture and the world-plane image registration.
\vspace{-5pt}
\subsection{Pedestrian Detection\label{sect:obj_dect}}
\subsubsection{Camera Dataset}
Footage is sourced openly via Transport for London sourced from the Open Roads initiative, known as JamCams \cite{jamcams}. A day of collection constitutes approximately 220,000 individual files of a total of 20-30GB, deleted upon processing in accordance with our data retention policy. The nature of monitoring public spaces means we cannot a priori request consent. The reductive resolution of footage collected from this source inhibits any capacity to personally identify an individual. Thus only their humanoid likeness is utilised for detection.
\vspace{-5pt}
\subsubsection{Object Detector}
In order to detect entities quickly enough to assist policy makers, we evaluate object detection models such as SSD \cite{liu2015ssd} and YOLO v3 \cite{redmon2018yolov3} to balance speed against accuracy. These are typically determined by architecture, model depth, input sizes, classification cardinality and execution environment. You Only Look Once (YOLO) \cite{redmon2015look} is a one-stage anchor-based object detector that is both fast and accurate.
YOLOv3 achieves an accuracy of 57.9 $AP_{50}$ in 51ms \cite{redmon2018yolov3}. Recently, a faster version named YOLOv4 \cite{bochkovskiy2020yolov4}, was released with a state-of-the-art accuracy than these alternative object detectors. Notably, YOLOv4 can be trained and used on conventional GPUs which allows for faster experimentation and fine-tuning on custom datasets. YOLOv4 improves performance and speed by 10\% and 12\% respectively \cite{bochkovskiy2020yolov4}.

We employ both YOLOv3 \& v4 in our experiments. Each were pretrained on Coco \cite{lin_microsoft_2014} dataset, a large-scale repository of objects belonging to 80 class labels.  Due to our objective, the classes of interest are limited to six labels: person, car, bus, motorbike, bicycle, and truck. We fine-tuned the model on six labels using joint datasets from COCO, MIO-TCD \cite{luo_mio_tcd_2018}, and a training set of custom manually labelled JamCam-specific set. A validation set was also partitioned from the manually labelled dataset for model evaluation. Results in the evaluation section documents the success of this fine-tuning to traffic camera footage.
\vspace{-5pt}
\begin{algorithm}[h!]
\caption{Group proximity frame tracking}\label{alg:groups}
\KwIn{Scene of localised detections, $S_L$}
\Parameter{Confidence threshold, $T_c$\newline Distance threshold, $T_d$}
\KwOut{Total detected groups, $G_n$\newline Max groups per-frame, $G_{\text{max}(n)}$\newline Min distance between groups, $G_{\text{min}(d)}$\newline Mean distance between groups, $G_{\overline{d}}$\newline Mean internal group size, $I_{\overline{n}}$\newline Mean internal group distance, $I_{\overline{d}}$}
\BlankLine
$\widetilde{S_L} \leftarrow S_{L,\text{conf}} \geq T_c$ \Comment*[r]{threshold confidence}
$I_l, I_d \leftarrow$ empty\;
 \ForEach{$f_L \in \widetilde{S_L}$ \Comment*[r]{locations per frame}}{ 
 $C \leftarrow$ empty\;
  \uIf{$|f_L| \leq 2$}{
  append($I_l$, $|f_L|$)\;
  \If{$|f_L| = 2$}{
  append($I_d$, Euclidean($f_{L_x}, f_{L_y}$))\;
  append($C$, Mean($f_{L_x}, f_{L_y}$))\;
  }
  }\Else{ 
  $E \leftarrow$ DelaunayEdges($f_L$)\;
  $D \leftarrow$ Euclidean($e_{v_x}$, $e_{v_y}$) $\forall e \in E$\;
  $\widetilde{E} \leftarrow E_d \leq T_d \forall d \in D$\Comment*[r]{threshold d.}
  $\boldsymbol{A} \leftarrow$ BuildCoordinateMatrix($\widetilde{E}$)\;
  \ForEach{$c \in \mathrm{ConnectedComponents}(\boldsymbol{A})$}{
    append($I_l$, $|c|$)\;
    append($I_d$, Mean($D_c$))\;
    append($C$, Mean($e_{c_v}$)\;
  }
  append($\widetilde{S_C}$, $C$)\Comment*[r]{group centres}
  }
}
\ForEach{$f_C \in \widetilde{S_C}$ \Comment*[r]{groups per frame}}{
    $E \leftarrow$ DelaunayEdges($f_C$)\;
    $L \leftarrow$ Euclidean($e_{v_x}$, $e_{v_y}$) $\forall e \in E$\;
    append($f_d$, Mean($L$))
 }
  $G_n \leftarrow |I_d|$\;
  $G_{\text{max}(n)} \leftarrow \text{max}(\text{max}(g) \hspace{0.4em} \forall g \in f_C)$\;
  $G_{\text{min}(d)} \leftarrow \text{min}(f_d)$\;
  $G_{\overline{d}} \leftarrow \text{Mean}(f_d)$\;
  $I_{\overline{n}} \leftarrow \text{Mean}(I_n)$\;
  $I_{\overline{d}} \leftarrow \text{Mean}(I_d)$\;
\end{algorithm}
\subsection{Scene Stability and Camera Drift\label{sect:sscd}}
\subsubsection{Similarity Indices}
Due to the extended duration of this project, it is necessary to include an evaluation of physical change in scene perspective or visible feed quality. Examples exhibited over time include intended adjustments made via motor-driven camera equipment, strong weather laterally progressively shifting direction, detachment from mounting hardware, and when local vegetation sprouts to inhibit visibility of the original scene. To mitigate and detect these issues, we construct a variation metric using past frame information to detect variations in the captured scene. Direct application of pixel-for-pixel Mean Square Error (MSE) is not suitable under the change of lighting conditions, and is too sensitive to minute pixel differences. The Structural Similarity Index Measure (SSIM) has been shown \cite{5596999} to aptly measure image distance via a kernel comparison approach.
$$\operatorname{SSIM}(x, y)=\frac{\left(2 \mu_{x} \mu_{y}+c_{1}\right)\left(2 \sigma_{x y}+c_{2}\right)}{\left(\mu_{x}^{2}+\mu_{y}^{2}+c_{1}\right)\left(\sigma_{x}^{2}+\sigma_{y}^{2}+c_{2}\right)}$$
where $\mu_x, \mu_y, \sigma^2_x, \sigma^2_y, \sigma_{xy}$ are $x$ and $y$ window average, variance, and covariance respectively; $c_{1,2}$ represent a weighted dynamic pixel range.

\begin{figure}[t]
\centering
\vspace{10pt}
\includegraphics[width=\columnwidth]{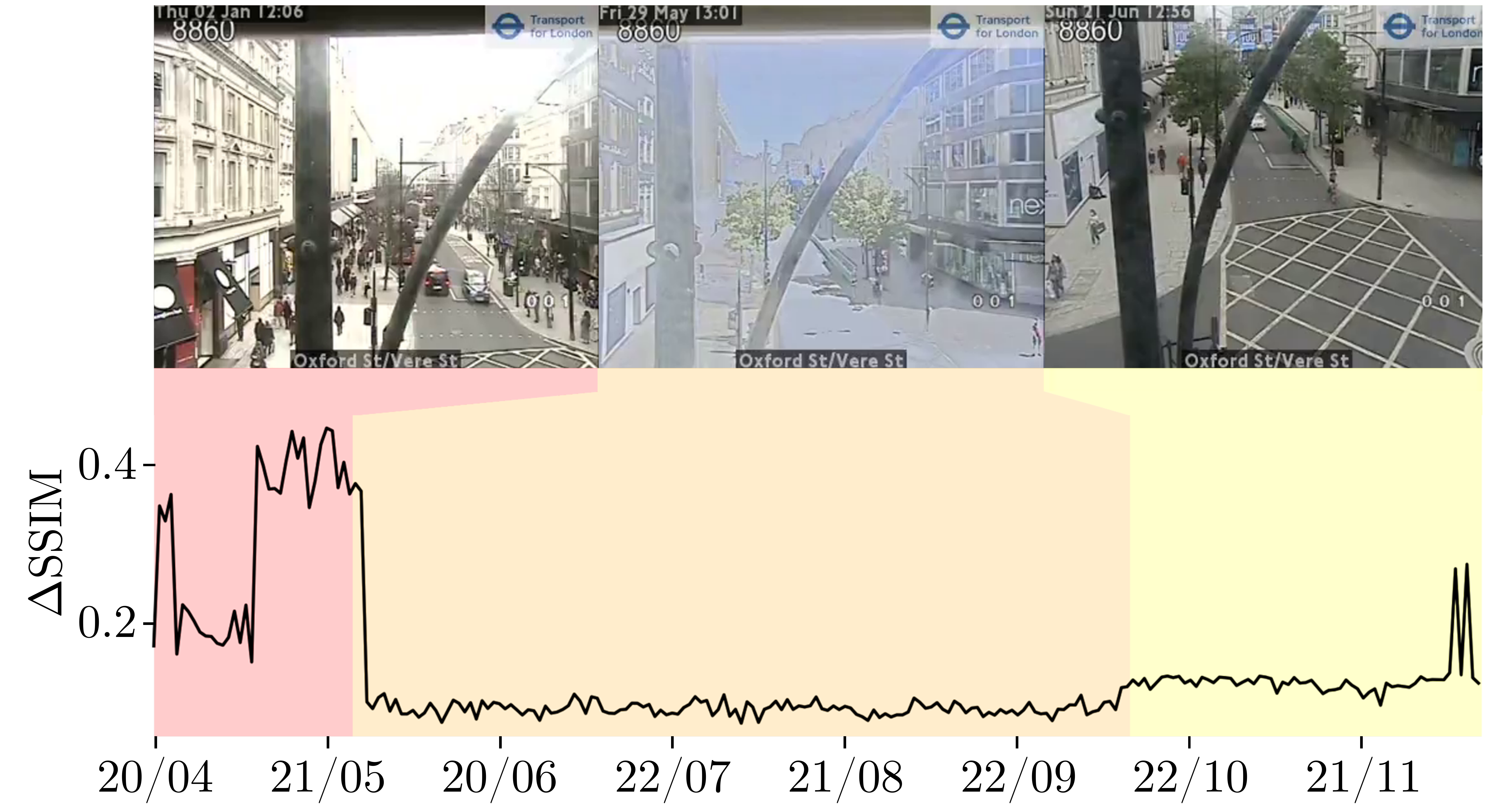}
\vspace{-8pt}
\caption{Change points detecting scene stability of Oxford St/Vere St. Each colour represents a detected change, example frames included and coloured respectively to indicate variation in camera quality and direction. \label{fig:camvar}}
\vspace{-5pt}
\end{figure}

\subsubsection{Application to scene imagery}
Numerous \textit{scene reference} periods were selected to measure SSIM for each camera: first known scene frame versus first hourly frame, one-week historical offset to first hourly frame, and mean of non-erroneous initial frames over seven days from initial data acquisition versus first daily at noon. It became clear that the final measure is most appropriate both for noise reduction and computational efficiency. Over time, this generates a univariate time series measuring scene variation. Sustained linear drift is less likely to negatively effect our pedestrian location estimation; however, it can be detected once the threshold is met. More importantly, a single major movement must be detected for subsequent alerting during the live experiment operation. This option however does deviate from other automated tasks, requiring the seven-day period to be adjusted to the new scene reference upon human intervention.

\subsubsection{Change Point Detection}
Detection of abrupt shifts in frame similarity over time is a task suitable to the unsupervised learning problem of \textit{change point detection}, the study of algorithms designed to find underlying change in time series \cite{burg2020evaluation}. An offline solution is still suitable, providing our large number of input feeds and necessity for appending daily measures. Upon evaluation, we determined Pruned Exact Linear Time (PELT) \cite{doi:10.1080/01621459.2012.737745} under a standard RBF kernel could accurately partition camera scene changes.
Under this measure, cameras of high variability are also excluded from later analysis.

\subsection{Group Detection\label{sect:group_detect}} In order to better describe social distancing efforts, we implement a group detection process (\cref{alg:groups}) and define seven metrics to describe a scene over time (\cref{tab:group_metrics}). Selecting groups from pedestrian detection locations are calculated by generating the Delaunay triangulation in the British National Grid (BNG) projection for pedestrians within individual frames per scene sample.

Each metric is calculated depending on two constants: detection confidence, $T_c$, the threshold required to include a detection, and a distance threshold, $T_d$, the maximum group diameter distance (metres). This task is conducted per frame, producing frame-level results: total number of groups, $G_n$; number of people within a group $I_l$; mean distance between individuals within group, $I_d$; and, group centres in BNG projection, $C$. Intermediate per-frame groups are determined by refactoring within threshold detection locations into a coordinate matrix from the Delaunay graph. Individual groups are then classified as connected components per \cite{Pearce2005AnIA}. Upon completion, each set of groups detected per frame supports an additional Delaunay triangulation, permitting final calculation of scene-level metrics: maximum number of groups per frame, $G_n$; minimum distance between groups; $G_{\text{max}(n)}$; and, mean distance between groups, $G_{\overline{d})}$.

We fix the detection confidence threshold to 0.7, meaning inferred detection certainties from object detection as pedestrian  below 70\% are excluded. A maximum interest area is defined as 6 metres between any two individuals. In practice, this task is expensive and distributed amongst many processing nodes using Python Dask parallelisation.

\subsection{Deployment\label{sect:deploy}}

\begin{figure}
    \centering
    \includegraphics[width=\columnwidth]{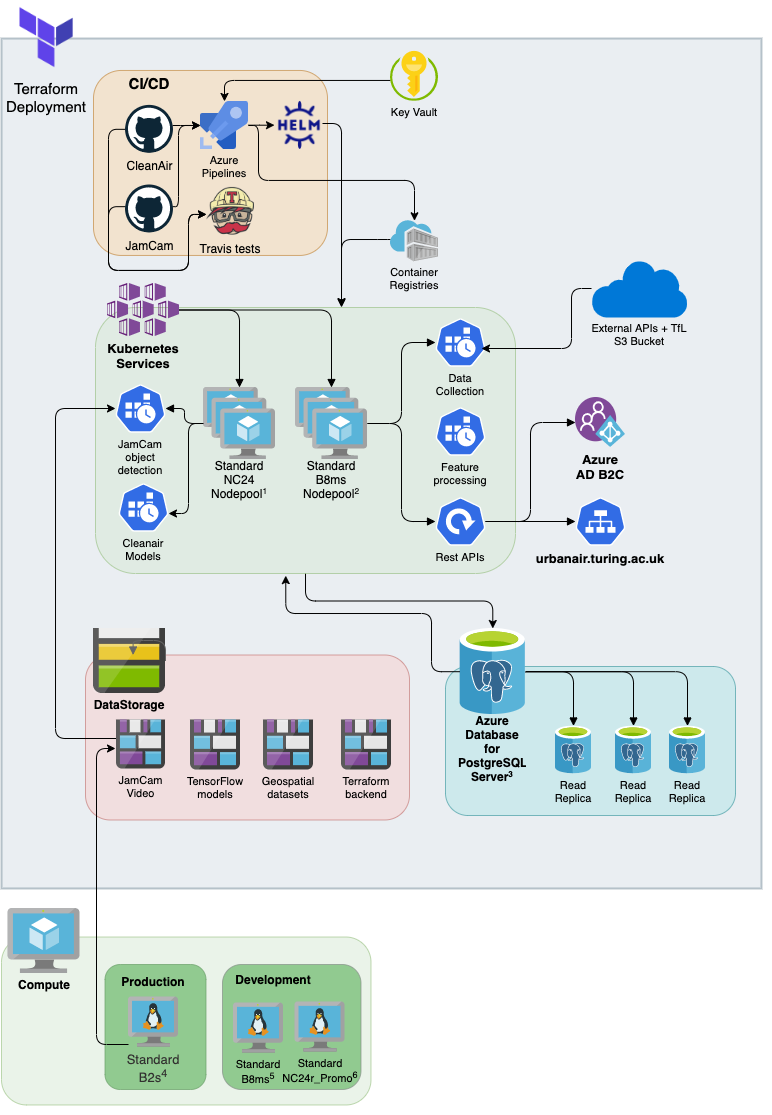}
    \vspace{-20pt}
    \caption{Development operations and platform architecture as deployed to Azure cloud services.\label{fig:deployment}}
    \vspace{-5pt}
\end{figure}

Deployment provisioning is controlled declaratively by Terraform, containing each component of the processing pipeline (\cref{fig:deployment}). Kubernetes manages two compute clusters: A GPU accelerated horizontally-scaled video processing pool, and a stability focused horizontally-scaled burstable CPU pool executing scheduled tasks and hosting API access points for direct data acquisition and service for control centre output.

\begin{table}
\small
\vspace{-11pt}
\begin{tabularx}{\columnwidth}{ll}\\\toprule  
\textbf{Metric} & \textbf{Definition}\\ \midrule
\textit{Individuals} & Total number of unique pedestrians \\ \arrayrulecolor{black!30}\midrule
\textit{No. groups} & Total number of unique groups \\  \arrayrulecolor{black!30}\midrule
\shortstack[l]{\textit{No. groups max.}\\\phantom{a}} & \shortstack[l]{Maximum number of exhibited\\groups in a given location.} \\  \arrayrulecolor{black!30}\midrule
\shortstack[l]{\textit{Outer group}\\\textit{min. distance}} & \shortstack[l]{Minimum distance exhibited between\\two groups in a given location.} \\  \arrayrulecolor{black!30}\midrule
\shortstack[l]{\textit{Outer group}\\\textit{mean distance}} & \shortstack[l]{Mean distance of all exhibited \\groups in a given location.} \\  \arrayrulecolor{black!30}\midrule
\shortstack[l]{\textit{Inner group}\\\textit{mean size}} & \shortstack[l]{Mean population within a group.\\\phantom{a}} \\ \arrayrulecolor{black!30}\midrule
\shortstack[l]{\textit{Inner group}\\\textit{mean. distance}} & \shortstack[l]{Mean of distance exhibited within\\a group.} \\ \arrayrulecolor{black!100}\bottomrule
\end{tabularx}
\caption{Group metrics calculated over a given sample of detection results.}\label{tab:group_metrics}
\vspace{-10pt}
\end{table} 
\section{Evaluation}\label{section_evaluation}

\subsection{Camera calibration}

\begin{table}
\centering
\small
\begin{tabularx}{\columnwidth}{@{\hskip3pt}l@{\hskip3pt}@{\hskip3pt}l@{\hskip3pt}@{\hskip3pt}l@{\hskip3pt}@{\hskip3pt}l@{\hskip3pt}@{\hskip3pt}l@{\hskip3pt}@{\hskip3pt}l@{\hskip3pt}@{\hskip3pt}l@{\hskip3pt}}
\toprule
\textbf{Dataset} &   {Person}  & {Bike} & {Car}  & {M.bike}  & {Bus}  & {Truck}\\ 
\midrule
\textit{Train}&&&&&& \\ \arrayrulecolor{black!30}\midrule
Coco 2017  &  262465   &   7113   &   43867   &   8725   &   6069   &   9973 \\
MIO-TCD    &    5760      &   1758   &   186767   &   1484   &   8443   &   54340 \\ \arrayrulecolor{black!100}\midrule
\textit{Validation}&&&&&& \\ \arrayrulecolor{black!30}\midrule
Coco 2017  &   11004  &   316   &   1932   &   371   &   285   &   415 \\
MIO-TCD &  1368  &   502   &   46730   &   353   &   2155   &   13694 \\
Jamcam  &  1233    &   106   &   7867   &   106   &   203   &   1982  \\ \arrayrulecolor{black!100}\bottomrule
\end{tabularx}
\caption{Training and validation samples per dataset.\label{tab: training_stats}}
\vspace{-15pt}
\end{table}

\subsubsection{World-plane estimation}
Uncertainty in the estimation of the vanishing line and extrinsic camera height arises due to imperfect camera effects eliminated in the assumptions and inaccurate automatic line extraction. The estimated errors in mapped world position, $dX, dY$, are evaluated for 3 randomly selected cameras manually calibrated beforehand using the total differential over all estimated parameters $p_i \in \{u_0,v_0,u_1,h\}$ assuming that the vehicle tracking $u,v$ are accurate at the point of evaluation. The average relative uncertainty in position mapping due to parameter estimation $|\frac{dX}{X}|$ is calculated to be 17.7\%, $\sigma$ = 7.9\%.

\subsubsection{Calibration to real-world reference}
Of 912 total cameras, 504 were selected for analysis within the Boroughs of Inner London with non-pedestrian motorway scenes predominately excluded. For this training subset, 3,298 manually labelled urban furniture anchors were employed for frame real-world calibration. During labelling, care was taken to maximise spacial coverage in each dimension, i.e. anchors were sparsely labelled to include the width and depth of the field of view. There are an average of 5.53 labels, $|F_s|$, per scene $s$, with a minimum of 4, $|F_s|\geq 4$, where few urban furniture could be identified. Given we are interested in the distance between individuals, the most appropriate error would be distance between known real-world locations and their pixel coordinates post transformation. The value of this comparison stems from the interest of comparing two individuals or groups within the scene. All possible lengths between all anchors, $N$, were calculated before optimisation. The error function was the mean squared error between these and the learned transformation results, $M$,

$$ \varepsilon = \text{MSE} \left( N, \sum^M_i{\sum^M_{j \ne i}{ \sqrt{i_{x,y}^2 - j_{x,y}^2} }} \right).$$

\begin{figure}[t]
\centering
\vspace{3pt}
\includegraphics[width=\columnwidth]{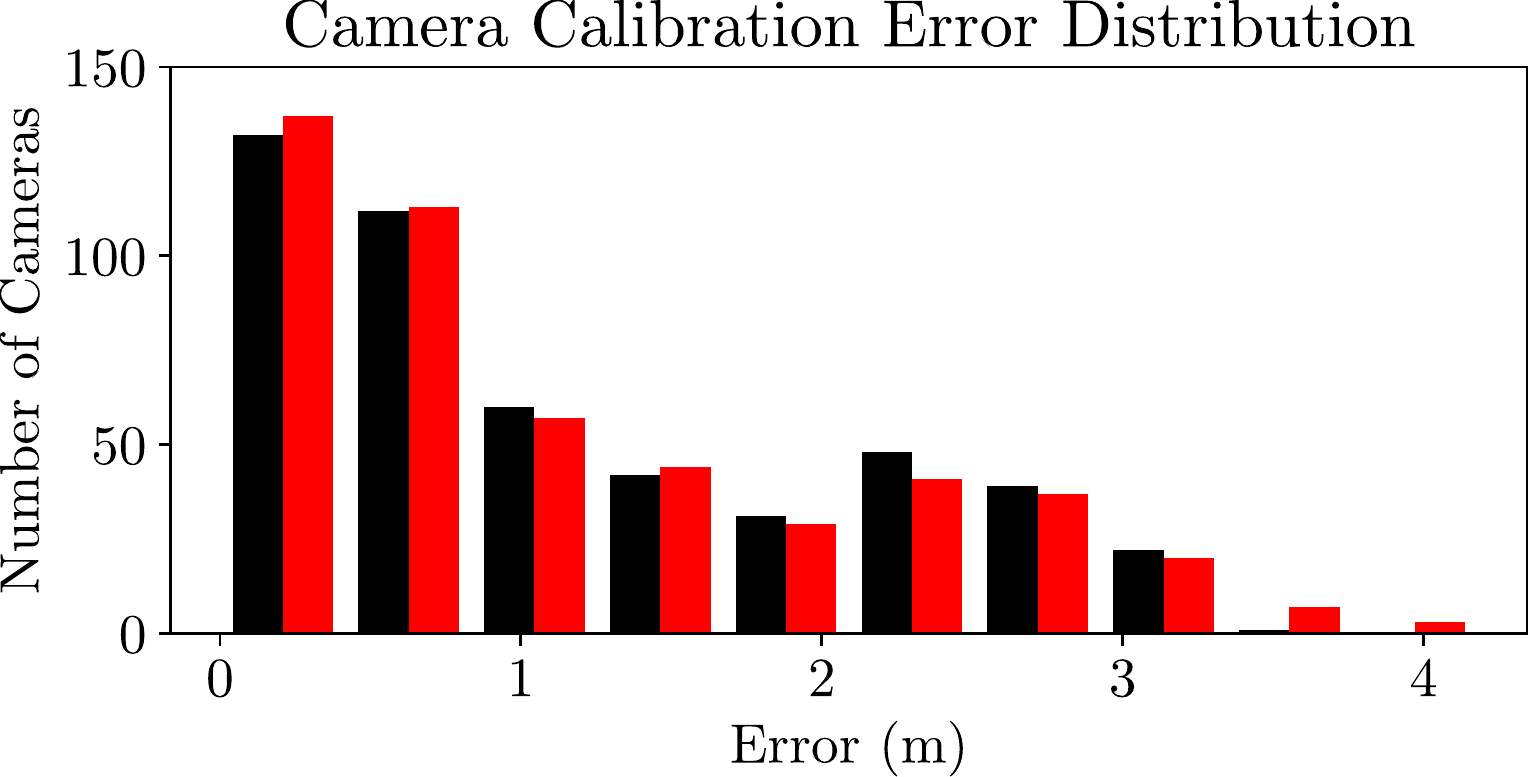}

\vspace{-10pt}
\caption{MSE distances of all possible distances from ground-truth anchors and transformation results. Full data are displayed in black, dropout validation results in red.\label{fig:camera_cal}}
\vspace{-5pt}
\end{figure}

This tests the complete calibration pipeline, from pixel coordinates in the image plane, to relative locations in the world plane, finally to the real-world distances between points. The median optimisation error was 0.8210 in BNS, meaning our model is able to locate an object in the image within 82.10cm. The distribution of this error is shown in \cref{fig:camera_cal}.

\subsubsection{Validation}

There does not exist a ground-truth dataset containing relative distances between people in the traffic camera frames around London. To validate this approach, we remove ground-truth anchors enforcing a reliance on fewer manually calibrated examples. For every scene $s$ with a set $F_s$ of urban furniture anchors, we remove exactly one anchor $a_s$ from $F_s$ randomly with uniform and independent probability. We train our model only using anchors in $F_s - \{a_s\}$. The validation test set contains all the out-of-sample removed anchors $a_s$ for each scene $s$.

The approach led to a mean relative distance error 83.43cm, a discrepancy of 1.33cm, with distribution displayed in red, \cref{fig:camera_cal}. This indicates that the training procedure marginally benefits from more labels and is resistant to changes in the input training data.

\subsection{Object detection} As preprocessing steps, we subset 6 labels from the Coco 2017 and MIO-TCD localization dataset. Unlike the Coco dataset, MIO-TCD localization dataset contains 11 labels with additional categories such as motorized vehicle, non-motorized vehicle, pickup truck, single unit truck, and work van, not found in the Coco 2017 dataset. For comparison, we collapse the different categories of trucks as \textit{truck} and remove labels regarding vehicle motorization. We produce a new collection of manually labelled entities specifically on frames from traffic cameras, using CVAT \cite{cvat}. The dataset contains 1142 frames and 11497 bounding boxes as shown in Table \ref{tab: training_stats}. For evaluation/validation, we compute the mean Average Precision (mAP) at IOU threshold of 0.5 over the Coco 2017, MIO-TCD, joint (Coco 2017 + MIO-TCD), and \textit{JamCam} datasets.

We fine-tune a pretrained YOLOv4 weights file on six labels from different training datasets using a batch size of 16, subdivsions of 4, image size of 416 and at least 7000 iterations on a Tesla V100-SXM3-32GB GPU. We train three different models on 1) Coco 2017 training data 2) MIO-TCD training data 3) Joint data containing random shuffle of Coco 2017 and MIO-TCD training data. Table \ref{tab: training_stats} shows the number of training data by labels. The validation data contains Coco 2017 validation data, MIO-TCD validation data and Jamcam data.

\begin{table}
\centering
\setlength{\tabcolsep}{3pt}
\small
\begin{tabular}{llcccc}
    \toprule
    {Train} & {Validation} & {mAP@0.50} & {Precision}  & {Recall}  & {F1-score} \\ \midrule
    \multirow{3}{*}{Coco}   & Coco  &   \textbf{67.55}   &   \textbf{0.73}  &   \textbf{0.70} &    \textbf{0.71} \\ 
                            & MIO-TCD  &   20.39   &   0.38 &    0.49   &  0.43   \\
                            & JamCam  &   41.64   &   0.62 &  0.59 &  0.60     \\ 
        \midrule
    \multirow{3}{*}{\shortstack{MIO\\-\\TCD}}    & Coco  &   14.24   &    0.35   &    0.30   &    0.30   \\ 
                                & MIO-TCD & \textbf{85.80}  &  \textbf{0.83}  &  \textbf{0.90}   &   \textbf{0.86} \\ 
                                & JamCam  &   35.12   &    0.75 & 0.45 & 0.57   \\ 
        \midrule
    \multirow{3}{*}{Joint}  & Coco & 64.56 & 0.71 & 0.69 & 0.70  \\ 
                            & MIO-TCD  &  \textbf{80.32} & \textbf{0.79} & \textbf{0.88} & \textbf{0.83} \\  
                            & JamCam  &  46.53   &   0.76 & 0.57 & 0.65   \\ 
    \bottomrule
\end{tabular}
\caption{Comparing models fine-tuned on the Coco 2017 dataset, MIO-TCD dataset, and joint training set using YOLOv4 architecture.\label{tab: eval_result}}
\vspace{-15pt}
\end{table}

The performances of the three models are shown in Table \ref{tab: eval_result}. On the Coco 2017 validation data, the model achieves a mean Average Precision (mAP@0.50) of 67.55\%. However, the model trained on Coco 2017 dataset perform poorly on MIO-TCD localization validation dataset with an mAP of 20.39\%. Likewise,  the performance of the model trained on MIO-TCD dataset reduces greatly from 85.80\% to 14.24\% when Coco 2017 dataset is used as the validation dataset. This behaviour might be as a result of the differences in the resolutions and weather conditions in the two datasets.  Performing a joint training creates a balance between the two datasets and increases the model's performance on the independent Jamcam dataset.

\section{Analysis}\label{section_results}

As of {September 23th 2021}, in the 18-month period the data collection pipeline has processed {19.31} terabytes of footage, for a total of {23,839,346,160} samples of all detection types, spanning all calibrated camera scenes. Of these, {9,453,327,651} were rejected either due to irrelevant camera positioning, or out of caution when recorded during a period of high camera variability.

We define ``Inner Bouroughs'' as the Statutory Inner London according to the London Government Act of 1963. The results for this section are generated for a period beginning from our collection date, 23rd March 2020.

\subsection{Scene Stability and Camera Drift}

\begin{figure*}[t]
    \centering
    \textbf{Inner London Borough Social Distancing Profiles}\\
    {\small April 2020 - April 2021}
    \includegraphics[width=\textwidth]{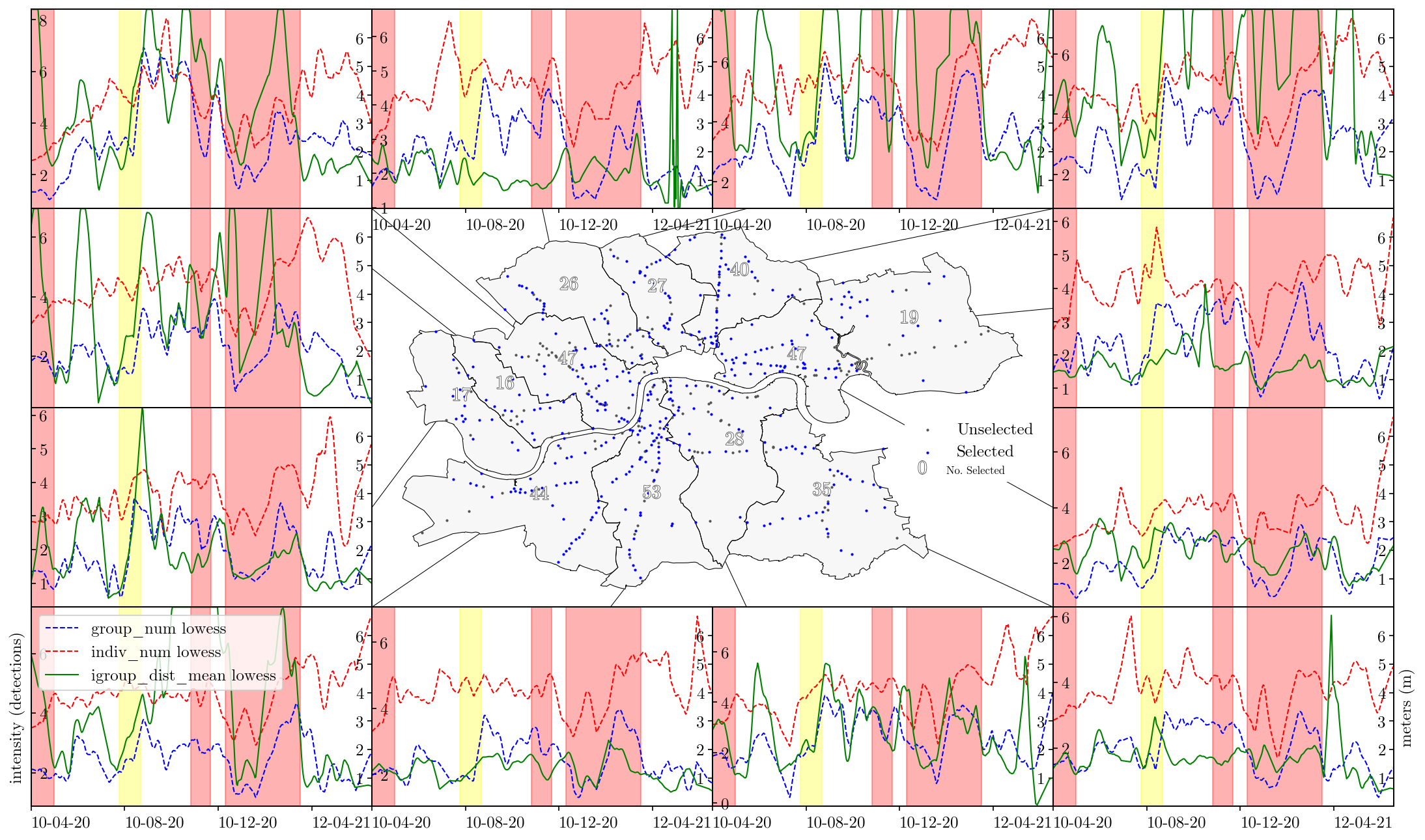} 
    \caption{Number of individuals, $I_{n}$, mean inner group physical distance, $I_{\overline{d}}$, \textit{outer} group social distance, $G_{\overline{d}}$ by inner borough. ``Lockdowns'' and ``Eat out to help out'' represented by red and yellow respectively. Points are representative of camera locations, \textit{selected} and \textit{unselected} in blue and grey respectively. \label{fig:macro} }
    \vspace{-10pt}
\end{figure*}

Change in SSIM between the \textit{reference scene} in the active feed, $\Delta \text{SSIM}$, is highly relevant to determining sample suitability. Instances of multiple change points promote manual intervention or total rejection, Figure \ref{fig:camvar} is an instance of changing stability, whereby the original perspective does not include both pedestrian crossings, then the sensor becomes damaged or over exposed over four months, only to then be positioned differently in October 2020.

Variance in camera positioning was noticeably larger in areas of high pedestrian activity, indicative of the active role Transport for London (TfL) has taken in monitoring pedestrian and vehicle traffic. As demonstrated by comparing the standard deviation of $\Delta \text{SSIM}$ between inner and non-inner boroughs over the aforementioned time frame results, $\sigma_{\text{inner}} = 0.0636$, $\sigma_{\text{outer}} = 0.0053$.

\subsection{Macro policy intervention}
Macro interventions within London are defined as either applicable national requirements determined by the central government or city-wide policies set forth by the Mayor's Office. For this example, inner boroughs are selected for their high camera availability for a 12-month subset of these data. Applying group detection per borough provides profiles visible in Figure \ref{fig:macro}. A timeline of intervention events \cite{restrictions_ts} are documented in Figure \ref{fig:macro_periods}. Each profile is smoothed under simple local regression \cite{doi:10.1080/01621459.1979.10481038}, taking 5\% closest points to ($x_i$,$y_i$), estimating $y_i$ under standard weighted linear regression.

There are two predominant results generalised across the City. A substantial increase in pedestrian frequency and social distance reduction during the ``Eat out to help out'' scheme between the 3rd to 30th August 2020. Additionally, activity during the second lockdown plummets whilst social distancing resumes a steady increase. Within the reduced restriction periods between these events, social distancing is seen to largely spike in most boroughs. Potentially indicative of successful public information campaigns and a willingness to maintain safe distancing conditions.

Within the Christmas period the initial social restriction measures rapidly stem the quantity of individuals and groups in all boroughs. Directly after the holiday passes, activity and distances rapidly grow until restrictions are relaxed leading to plummeting social distancing in almost all boroughs.
\subsection{Micro policy intervention}
\begin{figure*}[t]
    \centering
    \includegraphics[width=\textwidth]{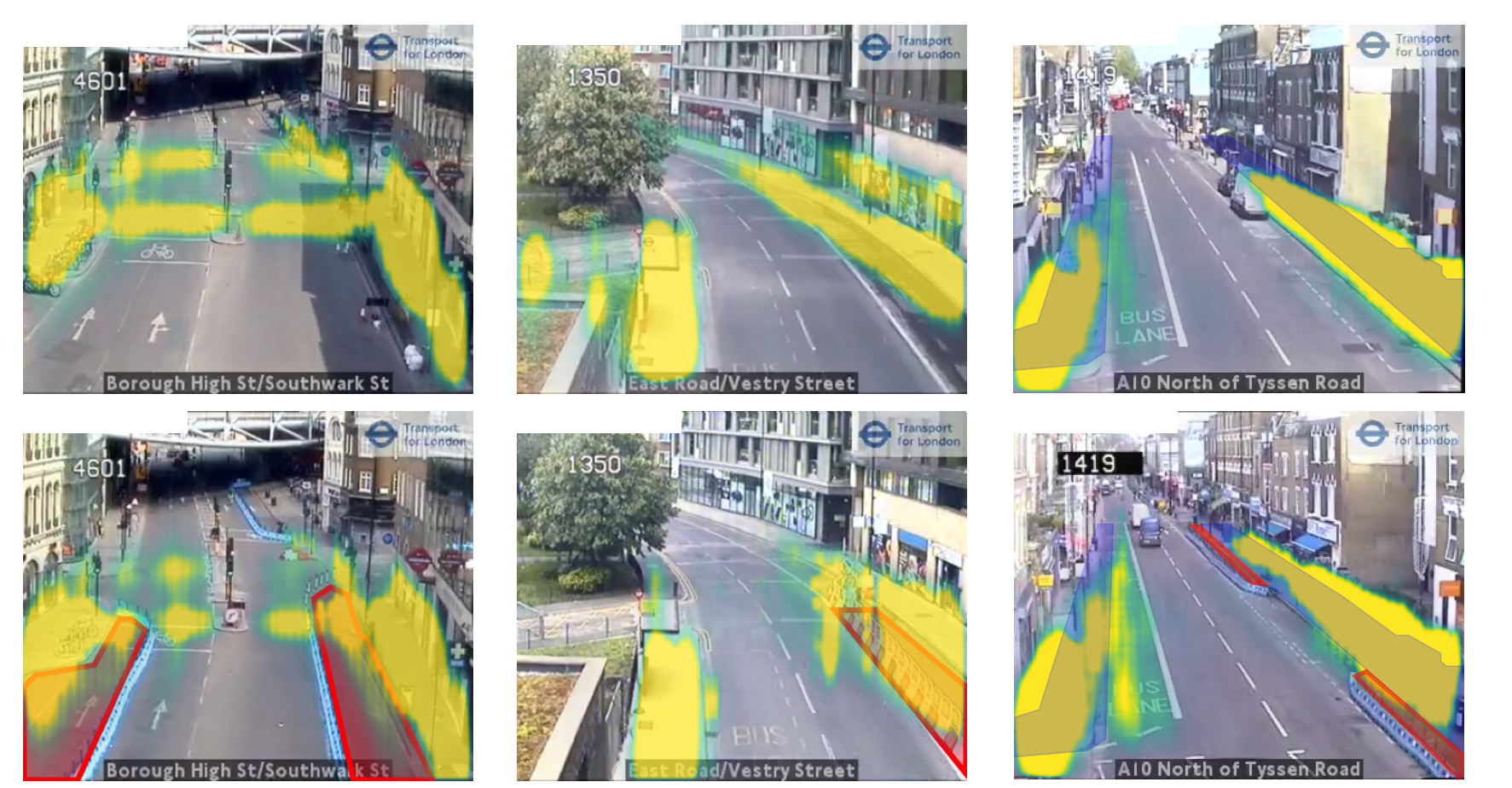}
    \caption{Before (top) and after (bottom) of three locations of pavement expansion interventions. Heat map of pedestrian footfall within calibrated pavement and extension (red) areas pre- and  post-bollard placement.}
    \label{fig:micro}
    \vspace{-15pt}
\end{figure*}
Micro interventions are limited within our dataset, as many COVID protocols cannot be captured on traffic cameras. There exist a number of smaller interventions in the form of pavement extensions. These expansions in pedestrian space include road reclamation in specific areas of high volume, such as near restaurants and public transport stations. For our analysis, we selected three stable scenes from distinct boroughs and filtered our social distance metrics before and after the intervention for comparison.

Locations \textit{Borough High St/Southward St}, \textit{East Road/Vestry St}, and \textit{A10 North of Tyssen Rd} (Figure \ref{fig:micro}) have 50.2, 43.0, and 153.1$\text{m}^2$ of pavement area within our digital twin scene. Post-expansion, each gained 40.2, 20.5, and 100.3$\text{m}^2$ of additional walking space. Before intervention these had mean estimated social distances of 1.33, 1.21, 0.73m, or as a ratio to area, 0.027, 0.028, 0.005 $\text{m}/ \text{m}^2$ respectively. Post barricade installation, estimated mean distance rose to 1.50, 1.25, 0.93m. This is indicative of a clear usage of this space, increased physical distancing, and effective policy intervention.

\section{Conclusion}\label{section_conclusion}
This work contributes a new social distancing monitoring platform, improves upon the accuracy of the state of the art detection model for in an urban domain, introduces a new camera perspective estimation method, provides physical spacing metrics into a viable historical context, and demonstrates how multiple machine learning techniques may benefit public health. According to the Greater London Authority, this tool enabled them to intervene quickly and identify where street spacing interventions were required. These interventions included moving bus stops, widening pavements and closing parking bays to create space which enabled social distancing. ``TfL says that it implemented over 700 such interventions at the height of the pandemic’s first wave, and that the Turing’s tool provided key data for those decisions \cite{impact1, impact2}.''

\begin{figure}[b!]
    \centering
    \includegraphics[width=\columnwidth]{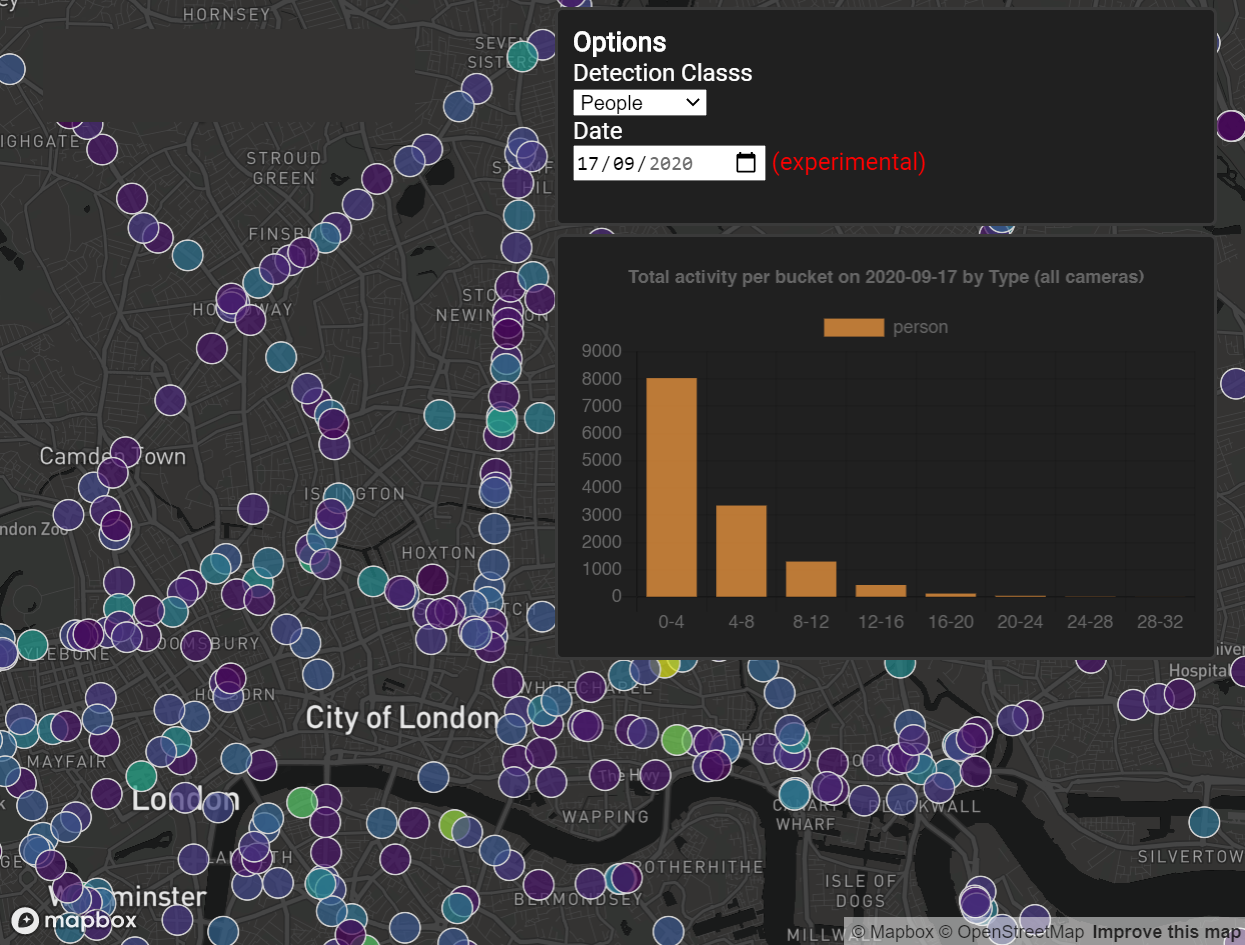}
    \caption{Control centre output, real-time interactive pan-London activity metrics as a web application convenient to stakeholders.\label{fig:control}}
    \vspace{-10pt}
\end{figure}

Combined with large-scale, inexpensive consumer distributed computing infrastructure, we provide an option for policy makers to receive a near real-time perspective of their impact via an online interface, \cref{fig:control}. Ongoing directions for this project include validating our early warning detection system, improving the digital twin overall accuracy, providing more "human-in-the-loop" recommendations with high ease of use for policy makers, and continuing to provide transparent and interrogatable examples of machine learning applications.

\ack{Funded by Lloyd’s Register Foundation programme on Data Centric Engineering and Warwick Impact Fund via the EPSRC Impact Acceleration Account. Further supported by the Greater London Authority, Transport for London, Microsoft, Department of Engineering at University of Cambridge and the Science and Technology Facilities Council. Camera location and footage are publicly available, the calibration data and a year of daily samples of CCTV footage are available in Zenodo, at \url{https://zenodo.org/record/6472731}. We would like to thank Sam Blakeman and James Brandreth for their help on multiple aspects of this work.}

\clearpage

\nocite{*}

\bibliographystyle{bcs_computer/compj}
\bibliography{bibliography}

\begin{thebibliography}{99}

\bibitem{gelenbe06}
Gelenbe, E. (2006) Analysis of automated auctions.
\newblock {\em ISCIS 2006, LNCS 4263},  Istanbul, Turkey,  1-3 November,  pp.
  1--12. Springer Verlag, Berlin.

\bibitem{gelenbe_inPress}
Gelenbe, E. (in press) Analysis of single and networked auctions.
\newblock Accepted for publication in \emph{ACM Trans. Internet Technology}.

\bibitem{Gelenbe-SoftwareViruses}
Gelenbe, E. (2007) Dealing with software viruses: A biological paradigm.
\newblock {\em Information Security Technical Report}, {\bf  12}, 242--250.

\bibitem{Gelenbe07_DiffModelPacketTravel}
Gelenbe, E. (2007) A diffusion model for packet travel time in a random
  multi-hop medium.
\newblock {\em ACM Trans. Sensor Netw.}, {\bf  3}, 10:1--10:19.

\bibitem{Gelenbe_06_Workshop_TravelDelay}
Gelenbe, E. (2006) Travel delay in a large wireless ad hoc network.
\newblock {\em 2nd Workshop on Spatial Stochastic Modeling of Wireless Networks
  (SpaSWiN 2006)},  Boston,  7 Apr.,  pp. 1--6. IEEE.

\bibitem{Gelenbe03_Sensible}
Gelenbe, E. (2003) Sensible decisions based on {QoS}.
\newblock {\em Computational Manage. Sci.}, {\bf  1}, 1--14.

\bibitem{zeithammer06}
Zeithammer, R. (2006) Forward-looking bidding in online auctions.
\newblock {\em J. Marketing Res.}, {\bf  43}, 462--476.

\bibitem{gagliano_etal_95}
Gagliano, R.~A., Fraser, M.~D., and Schaefer, M.~E. (1995) Auction allocation
  of computing resources.
\newblock {\em Commun. ACM}, {\bf  38}, 88--102.

\bibitem{maes99}
Maes, P., Guttman, R.~H., and Moukas, A.~G. (1999) Agents that buy and sell.
\newblock {\em Commun. ACM}, {\bf  42}, 81--91.

\bibitem{Dobson_etal_06}
Dobson, S., Denazis, S., Fern\'andez, A., Ga\"iti, D., Gelenbe, E., Massacci,
  F., Nixon, P., Saffre, F., Schmidt, N., and Zambonelli, F. (2006) A survey of
  autonomic communications.
\newblock {\em ACM Trans. Autonom. Adapt. Syst.}, {\bf  1}, 223--259.

\bibitem{Gelenbe_etal_Nagoya}
Gelenbe, E., Lent, R., Montuori, A., and Xu, Z. (2000) Towards networks with
  cognitive packets.
\newblock {\em Proc. of the Int. Conf. on Performance and {QoS} of Next
  Generation Networking},  Nagoya, Japan,  Nov.,  pp. 3--17. Springer, London.

\bibitem{Gelenbe_Gellman_Su_ISCC03}
Gelenbe, E., Gellman, M., and Su, P. (2003) Self-awareness and adaptivity for
  {QoS}.
\newblock {\em Proc. Eighth IEEE Int. Symp. on Computers and Communications
  (ISCC 2003)},  Kemer-Antalya, Turkey,  June,  pp. 3--9. IEEE Computer
  Society, Los Alamitos, California.

\bibitem{Gelenbe-Gellman-Lent-Liu04}
Gelenbe, E., Gellman, M., Lent, R., Liu, P., and Su, P. (2004) Autonomous smart
  routing for network {QoS}.
\newblock {\em Proc. Int. Conf. on Autonomic Computing (ICAC 2004)},  New York,
   17-18 May,  pp. 232--239. IEEE Computer Society, Los Alamitos, California.

\bibitem{Gelenbe_01_CompNetw_CPN}
Gelenbe, E., Lent, R., and Xu, Z. (2001) Measurement and performance of a
  cognitive packet network.
\newblock {\em Computer Networks}, {\bf  37}, 691--701.

\bibitem{Gelenbe_Mascots02}
Gelenbe, E., Lent, R., Montuori, A., and Xu, Z. (2002) Cognitive packet
  networks: {QoS} and performance.
\newblock {\em Proc. 10th IEEE Int. Symp. on Modelling, Analysis and Simulation
  of Computer and Telecommunications Systems (MASCOTS 2002)},  Ft. Worth,
  Texas,  11-16 Oct.,  pp. 3--12. IEEE Computer Society, Los Alamitos,
  California.

\bibitem{Gelenbe_Lent_Nunez_04}
Gelenbe, E., Lent, R., and Nunez, A. (2004) Self-aware networks and {QoS}.
\newblock {\em Proc. of the {IEEE}}, {\bf  92}, 1478--1489.

\bibitem{Gelenbe_Liu06_workshop}
Gelenbe, E., Liu, P., and Laine, J. (2006) Genetic algorithms for autonomic
  route discovery.
\newblock {\em Proc. IEEE Workshop on Distributed Intelligent Systems:
  Collective Intelligence and Its Applications (DIS 2006)},  Prague, Czech
  Republic,  15-16 June,  pp. 371--376. IEEE Computer Society, Los Alamitos,
  California.

\bibitem{Gelenbe_Liu06_GeneticAlgorithmsRouteDiscovery}
Gelenbe, E., Liu, P., and Laine, J. (2006) Genetic algorithms for route
  discovery.
\newblock {\em IEEE Trans. on Systems, Man and Cybernetics B}, {\bf  36},
  1247--1254.

\bibitem{Gelenbe-Lent_2004}
Gelenbe, E. and Lent, R. (2004) Power-aware ad hoc cognitive packet networks.
\newblock {\em Ad Hoc Networks}, {\bf  2}, 205--216.

\bibitem{gelenbe_sakellari_08}
Gelenbe, E., Sakellari, G., and D'Arienzo, M. (2008) Admission of {QoS} aware
  users in a smart network.
\newblock {\em ACM Trans. Autonom. Adapt. Syst.}, {\bf  3}, 4:1--4:28.

\bibitem{Gelenbe05_UsersServicesIntNetworks}
Gelenbe, E. (2005) Users and services in intelligent networks.
\newblock {\em Asian Internet Engineering Conf. (AINTEC 2005), LNCS 3837},
  Bangkok, Thailand,  13-15 December,  pp. 30--45. Springer Verlag, Berlin.

\bibitem{ferdinando_etal08}
Di~Ferdinando, A., Rosi, A., Zambonelli, F., Lent, R., and Gelenbe, E. (2008) A
  platform for pervasive combinatorial trading with opportunistic
  self-aggregation.
\newblock {\em IEEE Int. Symp. on A World of Wireless, Mobile and Multimedia
  Networks (WOWMOM 2008)},  Newport Beach, CA,  23-26 June,  pp. 1--6. IEEE.

\bibitem{Gelenbe_Loukas07_SelfAwareDoS}
Gelenbe, E. and Loukas, G. (2007) A self-aware approach to denial of service
  defence.
\newblock {\em Computer Networks}, {\bf  51}, 1299--1314.

\bibitem{Loukas_Oke_07}
Loukas, G. and Oke, G. (2007) A biologically inspired denial of service
  detector using the random neural network.
\newblock {\em Workshop on Socially and Biologically Inspired Wired and
  Wireless Networks (BIONETWORKS 2007)},  Pisa, Italy,  8-11 October,  pp.
  1--6. IEEE.

\bibitem{Loukas_Oke_LikelihoodRatios}
Loukas, G. and Oke, G. (2007) Likelihood ratios and recurrent random neural
  networks in detection of denial of service attacks.
\newblock {\em Proc. Int. Symp. on Performance Evaluation of Computer and
  Telecommunication Systems (SPECTS 2007)},  San Diego, California,  16-18
  July.

\bibitem{Oke_etal07}
Oke, G., Loukas, G., and Gelenbe, E. (2007) Detecting denial of service attacks
  with bayesian classifiers and the random neural network.
\newblock {\em Proc. IEEE Int. Conf. on Fuzzy Systems (FUZZ-IEEE 2007)},
  London, UK,  23-26 July,  pp. 1964--1969. IEEE.

\bibitem{krishna02}
Krishna, V. (2002) {\em Auction Theory},  1st. edition. Academic Press,
  California and London.

\bibitem{roth02}
Roth, A. and Ockenfels, A. (2002) Last-minute bidding and the rules for ending
  second-price auctions: Evidence from ebay and amazon auctions on the
  internet.
\newblock {\em Amer. Econ. Rev.}, {\bf  92}, 1093--1103.

\bibitem{ockenfels06}
Ockenfels, A. and Roth, A.~E. (2006) Late and multiple bidding in second price
  internet auctions: Theory and evidence concerning different rules for ending
  an auction.
\newblock {\em Games Econ. Behav.}, {\bf  55}, 297--320.

\bibitem{bajari03}
Bajari, P. and Hortacsu, A. (2003) The winner's curse, reserve prices, and
  endogenous entry: Empirical insights from ebay auctions.
\newblock {\em RAND J. Econ.}, {\bf  34}, 329--355.

\bibitem{popenda87}
Popenda, J. (1987) One expression for the solutions of second order difference
  equations.
\newblock {\em Proc. Amer. Math. Soc.}, {\bf  100}, 87--93.

\bibitem{mallik97}
Mallik, R. (1997) On the solution of a second order linear homogeneous
  difference equation with variable coefficients.
\newblock {\em J. Math. Anal. Appl.}, {\bf  215}, 32--47.

\bibitem{boese02}
Boese, F.~G. (2002) On ordinary difference equations with variable
  coefficients.
\newblock {\em J. Math. Anal. Appl.}, {\bf  273}, 378--408(31).

\bibitem{hillier64}
Hillier, F.~S., Conway, R.~W., and Maxwell, W.~L. (1964) A multiple server
  queueing model with state dependent service rate.
\newblock {\em J. Ind. Eng.}, {\bf  15}, 153--157.

\bibitem{david_rogers_jennings07}
David, E., Rogers, A., Jennings, N.~R., Schiff, J., Kraus, S., and Rothkopf,
  M.~H. (2007) Optimal design of english auctions with discrete bid levels.
\newblock {\em ACM Trans. Internet Technology}, {\bf  7}, 12:1--12:34.

\bibitem{gelenbe93}
Gelenbe, E. (1993) Learning in the recurrent random neural network.
\newblock {\em Neural Computation}, {\bf  5}, 154--164.

\bibitem{gelenbe_mitrani80}
Gelenbe, E. and Mitrani, I. (1980) {\em Analysis and Synthesis of Computer
  Systems}. Academic Press,  New York and London.

\bibitem{gelenbe_pujolle98}
Gelenbe, E. and Pujolle, G. (1998) {\em Introduction to Networks of Queues}. J.
  Wiley \& Sons,  Chichester.

\bibitem{guo02}
Guo, X. (2002) An optimal strategy for sellers in an online auction.
\newblock {\em ACM Trans. Internet Technology}, {\bf  2}, 1--13.

\bibitem{lucking07}
Lucking-Reiley, D., Bryan, D., Prasad, N., and Reeves, D. (2007) Pennies from
  ebay: the determinants of price in online auctions.
\newblock {\em J. Ind. Econ.}, {\bf  55}, 223--233.

\bibitem{medhi94}
Medhi, J. (1994) {\em Stochastic Processes}. J. Wiley \& Sons.

\bibitem{milgrom_weber82}
Milgrom, P.~R. and Weber, R.~J. (1982) A theory of auctions and competitive
  bidding.
\newblock {\em Econometrica}, {\bf  50}, 1089--1122.

\bibitem{orenRothkopf75}
Oren, S.~S. and Rothkopf, M.~H. (1975) Optimal bidding in sequential auctions.
\newblock {\em Oper. Res.}, {\bf  23}, 1080--1090.

\bibitem{rothkopfHarstad94}
Rothkopf, M.~H. and Harstad, R.~M. (1994) Modeling competitive bidding: a
  critical essay.
\newblock {\em Manage. Sci.}, {\bf  40}, 364--384.

\bibitem{shehory02}
Shehory, O. (2002) Optimal bidding in multiple concurrent auctions.
\newblock {\em Int. J. Coop. Inf. Syst.}, {\bf  11}, 315--327.

\end{thebibliography}


\begin{thebibliography}{99}

\bibitem{socdist_googletrends}
GoogleTrends (2021).
\newblock {Social Distancing Keyword Trend}.
\newblock
  \url{https://trends.google.com/trends/explore?date=2020-01-01%202021-09-13&q=social%20distancing}.
\newblock [Online; accessed 11-Sept-2021].

\bibitem{socdist_googlebooks}
GoogleBooks (2021).
\newblock {Social Distancing Keyword in Google Books}.
\newblock
  \url{https://www.google.com/search?q=social+distancing&source=lnt&tbs=cdr%3A1%2Ccd_min%3A0%2F0%2F0000%2Ccd_max%3A1%2F1%2F2020&tbm=bks}.
\newblock [Online; accessed 11-Sept-2021].

\bibitem{may_lockdown-type_2020}
May, T. (2020) Lockdown-type measures look effective against covid-19.
\newblock {\em BMJ}, {\bf  370}.
\newblock Publisher: British Medical Journal Publishing Group Section:
  Editorial.

\bibitem{gla_covid_mobility}
Authority, T. G.~L. (2021).
\newblock {Coronavirus (COVID-19) Mobility Report}.
\newblock
  \url{https://data.london.gov.uk/dataset/coronavirus-covid-19-mobility-report}.
\newblock [Online; accessed 16-Oct-2021].

\bibitem{ogl}
{HM Government} (2020).
\newblock Open {Government} {Licence}.
\newblock
  \url{http://www.nationalarchives.gov.uk/doc/open-government-licence/version/3/}.
\newblock [Online; accessed 11-Sept-2021].

\bibitem{hamelijnck2019multiresolution}
Hamelijnck, O., Damoulas, T., Wang, K., and Girolami, M. (2019).
\newblock Multi-resolution multi-task gaussian processes.

\bibitem{BIGAZZI2012538}
Bigazzi, A.~Y. and Figliozzi, M.~A. (2012) Congestion and emissions mitigation:
  A comparison of capacity, demand, and vehicle based strategies.
\newblock {\em Transportation Research Part D: Transport and Environment}, {\bf
   17}, 538--547.

\bibitem{turing_eag}
{The Alan Turing Institute} (2021).
\newblock {Ethics Advisory Group}.
\newblock
  \url{https://www.turing.ac.uk/research/data-ethics/ethics-advisory-group}.
\newblock [Online; accessed 16-Oct-2021].

\bibitem{k8s}
{Cloud Native Computing Foundation} (2022).
\newblock {Kubernetes}.
\newblock \url{https://kubernetes.io/}.
\newblock [Online; accessed 04-Feb-2022].

\bibitem{bochkovskiy2020yolov4}
Bochkovskiy, A., Wang, C.-Y., and Liao, H.-Y.~M. (2020) Yolov4: Optimal speed
  and accuracy of object detection.
\newblock {\em ArXiv}, {\bf  abs/2004.10934}.

\bibitem{yolov4github}
Redmon, J.~C., Bochkovskiy, A., Wang, C.-Y., and Liao, H.-Y.~M. (2022).
\newblock {YoloV4 implementation}.
\newblock \url{https://github.com/AlexeyAB/darknet}.
\newblock [Online; accessed 04-Feb-2022].

\bibitem{postgres}
{The PostgreSQL Global Development Group} (2022).
\newblock {PostgreSQL}.
\newblock \url{https://www.postgresql.org/}.
\newblock [Online; accessed 04-Feb-2022].

\bibitem{haycock2020expectationbased}
Haycock, C., Thorpe-Woods, E., Walsh, J., O'Hara, P., Giles, O., Dhir, N., and
  Damoulas, T. (2020) An expectation-based network scan statistic for a
  covid-19 early warning system.
\newblock {\em Machine Learning in Public Health workshop, Neural Information
  Processing Systems},  Anywhere, Earth,  11 December.

\bibitem{bubbles}
Guidance, B.~G. and Support (2021).
\newblock {UK Government Department of Health and Social Care guided ``Support
  Bubble''}.
\newblock
  \url{https://gov.uk/guidance/making-a-support-bubble-with-another-household}.
\newblock [Online; accessed 16-Oct-2021].

\bibitem{Dewitt2000ElementsOP}
Dewitt, B. and Wolf, P.~R. (2000) {\em {Elements of Photogrammetry (with
  Applications in GIS)}}. McGraw Hill,  New York, NY.

\bibitem{Zitov2003ImageRM}
Zitová, B. and Flusser, J. (2003) Image registration methods: a survey.
\newblock {\em Image and Vision Computing}, {\bf  21}, 977--1000.

\bibitem{caprile}
Caprile, B. and Torre, V. (1990) Using vanishing points for camera calibration.
\newblock {\em International Journal of Computer Vision}, {\bf  4}, 127--139.

\bibitem{faugeras}
Faugeras, O. (1993) {\em Three-dimensional computer vision: a geometric
  viewpoint}. MIT Press,  Cambridge, MA.

\bibitem{tsai}
Tsai, R. (1987) A versatile camera calibration technique for high-accuracy 3d
  machine vision metrology using off-the-shelf tv cameras and lenses.
\newblock {\em IEEE Journal on Robotics and Automation}, {\bf  3}, 323--344.

\bibitem{schoepflin}
{Schoepflin}, T.~N. and {Dailey}, D.~J. (2003) Dynamic camera calibration of
  roadside traffic management cameras for vehicle speed estimation.
\newblock {\em IEEE Transactions on Intelligent Transportation Systems}, {\bf
  4}, 90--98.

\bibitem{dubska}
Dubska, M., Herout, A., and Sochor, J. (2014) Automatic {Camera} {Calibration}
  for {Traffic} {Understanding}.
\newblock {\em Proceedings of the {British} {Machine} {Vision} {Conference}
  2014},  Nottingham,  pp. 42.1--42.12. British Machine Vision Association.

\bibitem{lai}
Lai, A. H.~S. and Yung, N. H.~C. (2000) Lane detection by orientation and
  length discrimination.
\newblock {\em IEEE Trans. Syst., Man, Cybern. B}, {\bf  30}, 539--548.

\bibitem{song}
Song, K.-T. and Tai, J.-C. (2006) Dynamic {Calibration} of
  {Pan}–{Tilt}–{Zoom} {Cameras} for {Traffic} {Monitoring}.
\newblock {\em IEEE Transactions on Systems, Man and Cybernetics, Part B
  (Cybernetics)}, {\bf  36}, 1091--1103.

\bibitem{dong}
Dong, R., Li, B., and Chen, Q.-m. (2009) An {Automatic} {Calibration} {Method}
  for {PTZ} {Camera} in {Expressway} {Monitoring} {System}.
\newblock {\em 2009 {WRI} {World} {Congress} on {Computer} {Science} and
  {Information} {Engineering}},  March,  pp. 636--640.

\bibitem{fung}
Fung, G. S.~K. (2003) Camera calibration from road lane markings.
\newblock {\em Optical Engineering}, {\bf  42}, 2967.

\bibitem{liebowitz}
Liebowitz, D. and Zisserman, A. (1998) Metric rectification for perspective
  images of planes.
\newblock {\em Proceedings. 1998 {IEEE} {Computer} {Society} {Conference} on
  {Computer} {Vision} and {Pattern} {Recognition} ({Cat}. {No}.{98CB36231})},
  June,  pp. 482--488.
\newblock ISSN: 1063-6919.

\bibitem{cathey}
Cathey, F. and Dailey, D. (2005) A novel technique to dynamically measure
  vehicle speed using uncalibrated roadway cameras.
\newblock {\em IEEE Proceedings. Intelligent Vehicles Symposium, 2005.},  pp.
  777--782. IEEE.

\bibitem{criminisi}
Criminisi, A. (2001) {\em Accurate {Visual} {Metrology} from {Single} and
  {Multiple} {Uncalibrated} {Images}}. Springer London,  London, UK.

\bibitem{schoepflin2}
Schoepflin, T.~N., Dailey, D.~J., et al. (2003) Algorithms for estimating mean
  vehicle speed using uncalibrated traffic management cameras. Technical
  report. Washington (State). Dept. of Transportation.

\bibitem{SimonyanZ14a}
Simonyan, K. and Zisserman, A. (2015) Very deep convolutional networks for
  large-scale image recognition.
\newblock In Bengio, Y. and LeCun, Y. (eds.), {\em 3rd International Conference
  on Learning Representations, {ICLR} 2015, San Diego, CA, USA, May 7-9, 2015,
  Conference Track Proceedings}.

\bibitem{7410521}
Xie, S. and Tu, Z. (2015) Holistically-nested edge detection.
\newblock {\em 2015 IEEE International Conference on Computer Vision (ICCV)},
  pp. 1395--1403.

\bibitem{8100105}
Liu, Y., Cheng, M.-M., Hu, X., Wang, K., and Bai, X. (2017) Richer
  convolutional features for edge detection.
\newblock {\em 2017 IEEE Conference on Computer Vision and Pattern Recognition
  (CVPR)},  pp. 5872--5881.

\bibitem{shapiro}
{Shapiro}, L. and {Stockman}, G. (2001) {\em Computer Vision}. Prentice House
  London,  London, UK.

\bibitem{zhang}
Zhang, Z., Tan, T., Huang, K., and Wang, Y. (2013) Practical {Camera}
  {Calibration} {From} {Moving} {Objects} for {Traffic} {Scene} {Surveillance}.
\newblock {\em IEEE Transactions on Circuits and Systems for Video Technology},
  {\bf  23}, 518--533.
\newblock Conference Name: IEEE Transactions on Circuits and Systems for Video
  Technology.

\bibitem{jamcams}
{Transport for London} (2020).
\newblock Our open data.
\newblock \url{https://www.tfl.gov.uk/info-for/open-data-users/our-open-data}.
\newblock [Online; accessed 11-Sept-2021].

\bibitem{liu2015ssd}
Liu, W., Anguelov, D., Erhan, D., Szegedy, C., Reed, S., Fu, C.-Y., and Berg,
  A.~C. (2016) Ssd: Single shot multibox detector.
\newblock In Leibe, B., Matas, J., Sebe, N., and Welling, M. (eds.), {\em
  Computer Vision -- ECCV 2016},  Cham,  pp. 21--37. Springer International
  Publishing.

\bibitem{redmon2018yolov3}
Redmon, J. and Farhadi, A. (2018) Yolov3: An incremental improvement.
\newblock {\em ArXiv}, {\bf  abs/1804.02767}.

\bibitem{redmon2015look}
Redmon, J., Divvala, S., Girshick, R., and Farhadi, A. (2016) You only look
  once: Unified, real-time object detection.
\newblock {\em 2016 IEEE Conference on Computer Vision and Pattern Recognition
  (CVPR)},  pp. 779--788.

\bibitem{lin_microsoft_2014}
Lin, T.-Y., Maire, M., Belongie, S., Hays, J., Perona, P., Ramanan, D.,
  Dollár, P., and Zitnick, C.~L. (2014) Microsoft {COCO}: {Common} {Objects}
  in {Context}.
\newblock In Fleet, D., Pajdla, T., Schiele, B., and Tuytelaars, T. (eds.),
  {\em Computer {Vision} – {ECCV} 2014},  Cham Lecture {Notes} in {Computer}
  {Science},  pp. 740--755. Springer International Publishing.

\bibitem{luo_mio_tcd_2018}
Luo, Z., Branchaud-Charron, F., Lemaire, C., Konrad, J., Li, S., Mishra, A.,
  Achkar, A., Eichel, J., and Jodoin, P.-M. (2018) {MIO}-{TCD}: {A} {New}
  {Benchmark} {Dataset} for {Vehicle} {Classification} and {Localization}.
\newblock {\em IEEE Transactions on Image Processing}, {\bf  27}, 5129--5141.
\newblock Conference Name: IEEE Transactions on Image Processing.

\bibitem{5596999}
{Horé}, A. and {Ziou}, D. (2010) Image quality metrics: Psnr vs. ssim.
\newblock {\em 2010 20th International Conference on Pattern Recognition},  pp.
  2366--2369.

\bibitem{burg2020evaluation}
van~den Burg, G. J.~J. and Williams, C. K.~I. (2020) An evaluation of change
  point detection algorithms.
\newblock {\em ArXiv}, {\bf  abs/2003.06222}.

\bibitem{doi:10.1080/01621459.2012.737745}
Killick, R., Fearnhead, P., and Eckley, I.~A. (2012) Optimal detection of
  changepoints with a linear computational cost.
\newblock {\em Journal of the American Statistical Association}, {\bf  107},
  1590--1598.

\bibitem{Pearce2005AnIA}
{Pearce, David J} (2005) {\em An improved algorithm for finding the strongly
  connected components of a directed graph}.
\newblock Victoria University.
\newblock Wellington, NZ.

\bibitem{cvat}
Sekachev, B., Manovich, N., Zhiltsov, M., Zhavoronkov, A., Kalinin, D., Hoff,
  B., TOsmanov, Kruchinin, D., Zankevich, A., DmitriySidnev, Markelov, M.,
  Johannes222, Chenuet, M., a andre, telenachos, Melnikov, A., Kim, J., Ilouz,
  L., Glazov, N., Priya4607, Tehrani, R., Jeong, S., Skubriev, V., Yonekura,
  S., vugia truong, zliang7, lizhming, and Truong, T. (2020).
\newblock opencv/cvat: v1.1.0.

\bibitem{restrictions_ts}
Authority, T. G.~L. (2021).
\newblock {GLA COVID-19 Restrictions Timeseries}.
\newblock
  \url{https://data.london.gov.uk/dataset/covid-19-restrictions-timeseries}.
\newblock [Online; accessed 16-Oct-2021].

\bibitem{doi:10.1080/01621459.1979.10481038}
Cleveland, W.~S. (1979) Robust locally weighted regression and smoothing
  scatterplots.
\newblock {\em Journal of the American Statistical Association}, {\bf  74},
  829--836.

\bibitem{impact1}
{James Lloyd} (2021).
\newblock {Helping London Navigate Lockdown Safely}.
\newblock
  \url{https://www.turing.ac.uk/research/impact-stories/helping-london-navigate-lockdown-safely}.
\newblock [Online; accessed 16-Oct-2021].

\bibitem{impact2}
Vryzakis, A., Snaith, B., and D'Addario, J. (2021).
\newblock {Project Odysseus - using existing infrastructure to tackle new
  problems}.
\newblock
  \url{https://theodi.org/article/project-odysseus-using-existing-infrastructure/}.
\newblock [Online; accessed 16-Nov-2021].

\bibitem{Haycock2020AnEN}
Haycock, C., Thorpe-Woods, E., Walsh, J., O'Hara, P., Giles, O., Dhir, N., and
  Damoulas, T. (2020) An expectation-based network scan statistic for a
  covid-19 early warning system.
\newblock {\em ArXiv}, {\bf  abs/2012.07574}.

\end{thebibliography}

\clearpage
\onecolumn
\begin{appendices}
\renewcommand\thefigure{\thesection.\arabic{figure}}    

\setcounter{figure}{0}    

\section{Lockdown periods}
\begin{figure}[h!]
    \centering
    \includegraphics[width=0.49\columnwidth]{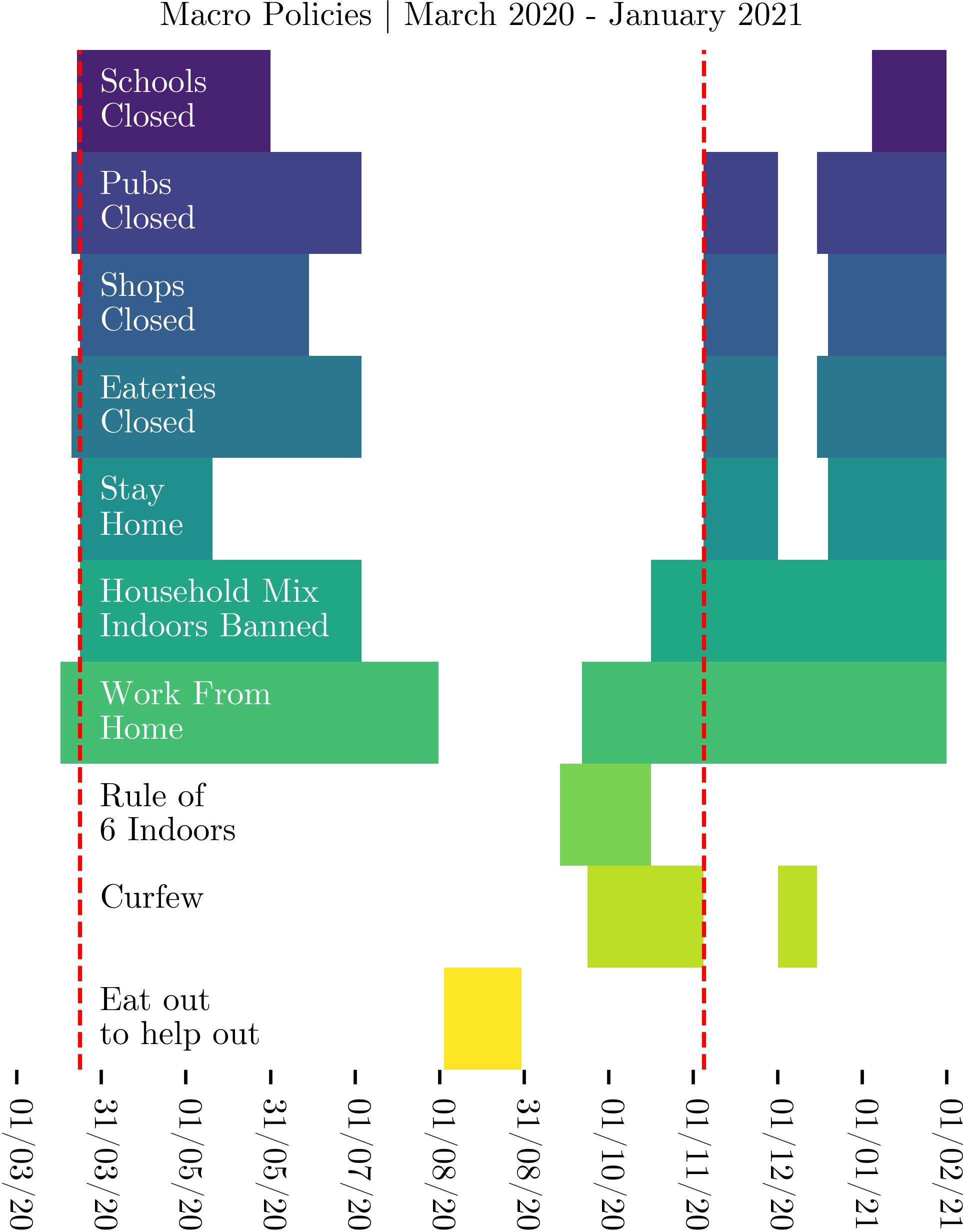}
    \caption{City-wide policy interventions. Red lines indicate lockdown start dates.\label{fig:macro_periods}}
\end{figure}

\newpage

\end{appendices}

\end{document}